\documentclass[a4paper,12pt]{article}

\begin{document}

\title{Antisymmetric solitons and their interactions in strongly
dispersion-managed fiber-optic systems}
\author{Bao-Feng Feng \\
Department of Mathematics, the University of Texas-Pan American, \\
1201 W. University Drive, Edinburg, Texas 78541-2999, U.S.A.\\
\and Boris A. Malomed \\
Department of Interdisciplinary Studies, Faculty of Engineering, \\
Tel Aviv University, Tel Aviv 69978, Israel}
\maketitle

\begin{center}
{\large \textbf{Abstract}}
\end{center}

By means of the variational approximation (VA), a system of ordinary
differential equations (ODEs) is derived to describe the propagation of
antisymmetric solitons in a multi-channel (WDM) optical fiber link subject
to strong dispersion management. Results are reported for a prototypical
model including two channels. Using the VA technique, conditions for stable
propagation of the antisymmetric dispersion-managed (ASDM) solitons in one
channel are found, and complete and incomplete collisions between the
solitons belonging to the different channels are investigated. In
particular, it is shown that formation of a bound inter-channel state of two
ASDM solitons is possible under certain conditions (but may be easily
avoided). The VA predictions for the single- and two-channel systems are
compared with direct simulations of the underlying partial differential
equations. In most cases, the agreement is very good, but in some cases
(very closely spaced channels) the collision may destroy the ASDM solitons.
The timing-jitter suppression factor (JSF) for the ASDM soliton in one
channel, and the crosstalk timing jitter induced by collision between the
solitons belonging to the different channels are also estimated
analytically. In particular, the JSF for the ASDM soliton may be much
larger than for its fundamental-soliton counterpart in the same system.

\section{Introduction}

The potential offered by the use of the dispersion management (DM), i.e.,
periodic compensation of the group-velocity dispersion in a long fiber-optic
telecommunication link, for the improvement of data transmission by soliton
streams, is well known \cite{Mecozzi}. The work in this direction,
especially aimed at the application of DM to multi-channel systems based on
the wavelength-division-multiplexing (WDM) technique, continues, see, e.g.,
a recent experimental report \cite{Japan-solitons}. The interest to the
topic has been recently bolstered by the development of the concept of the
differential phase-shift keying (instead of the traditional on-off code),
which helps to resolve some problems \cite{Mollenauer}, and by the launch
(in Australia) of the first commercial DM-soliton-based commercial
fiber-optic telecommunications link.

Basic properties of the fundamental solitons in DM systems have been studied
in detail by means of various analytical and numerical methods (some
references are in given in more particular contexts below). It is known
that, alongside the fundamental solitons which have a symmetric profile in
the temporal domain, \textit{antisymmetric} solitons are also possible in
the DM fiber-optic links \cite{Pare'} [in contrast to the uniform nonlinear
optical fibers, described by the nonlinear Schr\"{o}dinger (NLS) equation,
that does not give rise to antisymmetric solutions]. In the general case,
the asymmetric DM (ASDM) solitons may be unstable against parity-breaking
(symmetric) perturbations, but, nevertheless, in many respects they behave
as fairly robust pulses, that is why they are of interest to applications 
\cite{first-link}. Besides that, they are interesting dynamical objects in
their own right -- in particular, because they are related to the so-called
\textquotedblleft twisted localized modes\textquotedblright\ (TLM; alias
\textquotedblleft dark-in-bright solitons\textquotedblright ) in
Bose-Einstein condensates{\ (BECs)} loaded in a periodic potential ( optical
lattice) \cite{TLM}, and recently found odd solitons in BECs subjected to
the Feshbach-resonance management (time-periodic change of the sign of the
nonlinearity constant, under the action of external ac magnetic field) 
\cite{Feshbach}. Those solitons, in turn, were found following the pattern of
earlier found TLMs in the discrete NLS equation \cite{DNLS}. It is relevant
to mention that the TLM pulses in Bose-Einstein condensates are completely
stable objects, including full stability against parity-violating
perturbations \cite{TLM}.

The study of the ASDM solitons in single- and two-channel systems is the
subject of this work. First, we aim to develop the variational approximation
(VA) for the antisymmetric solitons in the single-channel DM model, in order
to predict conditions for stable transmission of these solitons in the long
DM fiber link. Then, we extend the VA for the case of interactions
(collision) between the ASDM solitons in the two-channel system. These
analytical results are presented in Section 2, and in Section 3 they are
verified versus direct numerical simulations. We infer that, in most cases,
the VA predictions are quite accurate, except for a case of two very close
channels, when collision may completely destroy the solitons. In Section 4,
we produce analytical estimates for the intra-channel jitter-suppression
factor (JSF), and for the crosstalk jitter induced by complete and
incomplete collisions between the ASDM solitons belonging to different
channels. A noteworthy result is that, for the antisymmetric solitons, the
JSF may be much larger (by a factor in excess of $10$) than for their
fundamental counterparts. The paper is concluded by Section 5.

\section{The analytical approach}

\subsection{The models}

We take the propagation equation for the DM transmission line in the
following standard normalized form (see, e.g., Refs. \cite{Anders,BorisVM}): 
\begin{equation}
2i u_{z}+D(z)u_{\tau \tau }+\epsilon (D_{0}u_{\tau \tau }+2|u|^{2}u)=0,
\label{1-ch}
\end{equation}
where $u(z,\tau )$ is the envelope of the electromagnetic field, $z$ and 
$\tau $ are the propagation distance along the fiber and the retarded time,
respectively, and $D(z)$ is the local dispersion coefficient, which is a
periodic function with a period $L_{\mathrm{map}}\equiv L_{1}+L_{2}$: 
\begin{equation}
D(z)\,=\,\left\{ 
\begin{array}{cl}
D_{1}{\,,} & 0<\mathrm{mod}(z,L_{\mathrm{map}})<L_{1}\,, \\ 
D_{2}{\,,} & L_{1}<\mathrm{mod}(z,L_{\mathrm{map}})<L_{1}+L_{2}~.
\end{array}
\right.  \label{map}
\end{equation}
The map (\ref{map}) is subject to the \textit{dispersion-compensation
condition,} $D_{1}L_{1}+D_{2}L_{2}=0$, and its parameters may be rescaled to
satisfy the following normalizations \cite{we2}, 
\begin{equation}
L_{1}+L_{2}=1,\;|D_{1}|L_{1}=|D_{2}|L_{2}=1.  \label{scaling}
\end{equation}

The small parameter $\epsilon $ in Eq. (\ref{1-ch}) is the ratio of the
local dispersion length to the nonlinear length, which measures the weakness
of the nonlinearity. The coefficient $D_{0}$ is the path-average dispersion
(PAD); its positive, zero, or negative values correspond, respectively, to
the anomalous, zero and the normal average dispersion, respectively. The
form of Eq. (\ref{1-ch}) implies that the nonlinearity and PAD are weak
factors at the same order of smallness.

A two-channel system for the fields $u(z,\tau )$ and $v(z,\tau )$
propagating in the same core obeys coupled nonlinear Schr\"{o}dinger
equations, 
\begin{equation}
2i\left( u_{z}+\overline{c}u_{\tau }\right) +D(z)u_{\tau \tau }+\epsilon
\left[ \overline{D}_{u}u_{\tau \tau }+2\left( \left\vert u\right\vert
^{2}+2\left\vert v\right\vert ^{2}\right) u\right] =0,  \label{2-cha}
\end{equation}
\begin{equation}
2i\left( v_{z}-\overline{c}v_{\tau }\right) +D(z)v_{\tau \tau }+\epsilon
\left[ \overline{D}_{v}v_{\tau \tau }+2\left( \left\vert v\right\vert
^{2}+2\left\vert u\right\vert ^{2}\right) v\right] =0,  \label{2-chb}
\end{equation}
where $2\overline{c}$ is the inverse group velocity difference between the
channels, $D(z)$ is the same periodic map as in Eq. (\ref{map}), while the
two PAD coefficients $\overline{D}_{u,v}$ may be different. Nonlinear terms
in Eqs. (1) and (2) represent the self-phase modulation (SPM) and
cross-phase modulation (XPM), which are induced by the Kerr effect 
\cite{Agrawal}. Equations (\ref{2-cha}), (\ref{2-chb}) conserve two channel
energies, 
\begin{equation}
E_{u}\equiv \sqrt{2/\pi }\int_{-\infty }^{+\infty }|u|^{2}d\tau
,\,E_{v}\equiv \sqrt{2/\pi }\int_{-\infty }^{+\infty }|v|^{2}d\tau 
\label{E}
\end{equation}
(the factor $\sqrt{2/\pi }$ was added to the definitions for convenience,
see below), as well the Hamiltonian and field momentum.

It is straightforward to establish relations between the normalized
variables and parameters used in the above equations, and their physical
counterparts. In particular, if $z=1$ and $\tau =1$ correspond to the
typical values, $50$ km and $10$ ps, respectively, which frequently play the
role of the length ant time units in fiber-optic telecommunications, then 
$D=1$ corresponds to the actual value of the dispersion coefficient 
$2$ ps$^{2}$/km (in uniform links, this value is typical for dispersion-shifted
fibers), and $\overline{c}=1$ corresponds to $0.2$ ps/km. With the
above-mentioned value of the dispersion coefficient, $2$ ps$^{2}$/km, the
inverse-group-velocity difference of $0.2$ ps/km between the channels
implies the wavelength separation $\Delta \lambda \approx 0.15$ nm between
them, which corresponds to the case of dense WDM.

The single-channel model (\ref{1-ch}) was used for the derivation of
conditions for stable transmission of fundamental DM solitons 
\cite{we2,DMsta}, and for the study of higher-order DM pulses based on the
Hermite-Gaussian functions \cite{Lakoba}. The two-channel model 
(\ref{2-cha}), (\ref{2-chb}) was used to investigate
inter-channel collisions between
fundamental pulses \cite{KMY}. Under certain conditions, the two-channel
model can also predict formation of bound states between two fundamental
solitons belonging to different channels, which was investigated in Ref. 
\cite{we}.

\subsection{The variational approximation for antisymmetric solitons}

The Hermite-Gaussian set of functions can be used to describe the
propagation of pulses of a general shape in the strong-DM regime, the
fundamental soliton of the Gaussian form being the first term in the set 
\cite{Lakoba}. The antisymmetric DM (ASDM) soliton corresponds to the second
function belonging to the set, so that it can be approximated by the
following variational ansatz: 
\begin{equation}
u(z,\tau )=A\tau \exp \left( -\frac{\tau ^{2}}{W^{2}}+ib{\tau }^{2}+i\phi
\right) ~.  \label{az1}
\end{equation}
Here, $A$, $W$, $b$ and $\phi $ represent the amplitude, width, chirp and
phase of the pulse, respectively, and they are allowed to be functions of 
$z$. The pulse is called the antisymmetric soliton because $\left| u(z,\tau
)\right| $ is an odd function of $\tau $. The energy of the pulse 
(\ref{az1}), calculated according to the definition (\ref{E}), 
is $E=A^{2}W^{3}/4$.

The antisymmetric soliton may also be represented in an alternative form, 
\begin{equation}
u(z,\tau )=A\tau \exp \left( -\frac{\tau ^{2}}{\tau {_{0}}^{2}+2i\Delta }
+i\phi \right) ,  \label{az2}
\end{equation}
where $\Delta (z)\equiv \int_{0}^{z}D\left( z^{\prime }\right) dz^{\prime
}+\Delta _{0}$ is the accumulated dispersion, and $\tau _{0}$ is the minimum
width of the pulse over the DM period. They are related to the parameters of 
$W$ and $b$ from the ansatz (\ref{az1}), 
\begin{equation}
W=\frac{\sqrt{\tau {_{0}}^{4}+4\Delta ^{2}}}{\tau _{0}},\qquad b=\frac{
2\Delta }{\tau {_{0}}^{4}+4\Delta ^{2}}.  \label{newpar}
\end{equation}
Following the commonly adopted definition \cite{Anders}, below we will be
using, instead of $\tau _{0}$, the so-called DM strength, 
\[
S\equiv 1.443/\tau _{0}^{2}. 
\]

The Lagrangian of the system (\ref{2-cha}), (\ref{2-chb}) is 
$L=\int_{-\infty }^{+\infty }\mathcal{L}dt$, with the Lagrangian density 
\begin{eqnarray}
{\mathcal{L}} &=&\frac{i}{2}\left[ (u_{z}u^{\ast }-uu_{z}^{\ast
}+v_{z}v^{\ast }-vv_{z}^{\ast })+\overline{c}\left( u_{\tau }u^{\ast
}-uu_{\tau }^{\ast }-v_{\tau }v^{\ast }+vv_{\tau }^{\ast }\right) \right]  
\nonumber \\
&&-\frac{1}{2}D(z)\left( |u_{\tau }|^{2}+|v_{\tau }|^{2}\right) -\frac{
\epsilon }{2}\left( \overline{D}_{u}|u_{\tau }|^{2}+\overline{D}_{v}|v_{\tau
}|^{2}\right)   \nonumber \\
&&+\frac{\epsilon }{2}\left( |u|^{4}+|v|^{4}+4|u|^{2}|v|^{2}\right) .
\label{lagden}
\end{eqnarray}
Applying the known technique of the VA for pulses of the Gaussian type \cite
{BorisVM}, and skipping routine technical details, we obtain the following
system of ordinary differential equations (ODEs) which govern the evolution
of parameters of the ansatz (\ref{az1}) in the single-channel system (with 
$\overline{c}=0$): 
\begin{eqnarray}
\frac{d\,E}{d\,z} &=&0~,  \label{conservation} \\
\frac{d\,W}{d\,z} &=&2\left[ D(z)+\epsilon D_{0}\right] \,b\,W~,  \label{dW}
\\
\frac{d\,b}{d\,z} &=&2\left[ D(z)+\epsilon D_{0}\right] \left( \frac{1}{W^{4}
}-b^{2}\right) -\frac{\sqrt{2}}{8W^{3}}\epsilon E.  \label{db}
\end{eqnarray}
where Eq. (\ref{conservation}) simply means that the energy is conserved. 
From Eqs. (\ref{newpar}), the evolution
ODEs for $\Delta _{0}$ and $\tau _{0}$ can also be derived: 
\begin{eqnarray}
\frac{d\,\tau _{0}}{d\,z} &=&\frac{\sqrt{2}}{16}\frac{\epsilon E\tau
_{0}\Delta }{W^{3}}~,  \label{dtau} \\
\frac{d\,\Delta _{0}}{d\,z} &=&\epsilon D_{0}+\frac{\sqrt{2}}{16}\frac{
\epsilon E(4\Delta ^{2}-\tau _{0}^{4})}{W^{3}}.  \label{ddel}
\end{eqnarray}

Conditions for stationary propagation of the ASDM soliton can be obtained in
the same way as it was done for the fundamental DM soliton in Ref. 
\cite{we2}. To this end, one should demand that the pulse's 
amplitude and width return
to the original values after passing one DM period, i.e., $\tau _{0}(z)=\tau
_{0}(z+1)$ and $\Delta _{0}(z)=\Delta _{0}(z+1)$. In the first-order
approximation (which implies that the parameters $\tau _{0}$ and $\Delta
_{0} $ suffer a small variation within one period), these conditions amount
to 
\begin{equation}
\int_{0}^{1}\frac{d\,\tau _{0}}{dz}d\,z=\int_{0}^{1}\frac{d\,\Delta _{0}}{dz}
d\,z=0.  \label{balance}
\end{equation}
Substituting Eqs. (\ref{dtau}), (\ref{ddel}) into Eqs. (\ref{balance}), and
evaluating some integrals explicitly, we obtain 
\begin{eqnarray}
\Delta _{0} &=&-1/2,  \label{del} \\
D_{0} &=&-\frac{\sqrt{2}}{16}E{\tau _{0}}^{3}\left[ \ln \left( \sqrt{1+{\tau
_{0}}^{-4}}+{\tau _{0}}^{-2}\right) -2\left( {\tau _{0}}^{4}+1\right)
^{-1/2} \right] .  \label{tau}
\end{eqnarray}
Note that the simple result (\ref{del}) is exactly the same as for the
fundamental solitons \cite{we2}, which implies that the pulse has zero chirp
at the midpoint of each fiber segment. Condition (\ref{tau}) is also similar
to the corresponding condition for the fundamental solitons, as derived in
Ref. \cite{we2}, only differing by a factor of $1/4$.

Straightforward extension of the VA-based analysis performed in Ref. 
\cite{we2} for the fundamental DM solitons, we arrive at the following
conclusions for their antisymmetric counterparts:

\begin{enumerate}
\item stable ASDM solitons exist at zero PAD if $S\approx 4.79$.

\item stable ASDM solitons exist at anomalous PAD if $S<4.79$.

\item stable ASDM solitons exist at normal PAD if $4.79<S<9.75$ and $| 
\overline{D}|/E\leq 0.0032$.
\end{enumerate}

Before proceeding further, we make several remarks. First, the stable
antisymmetric soliton is possible at both the anomalous PAD and normal PAD,
as well as when the PAD is zero. Second, the detailed VA analysis shows that
the energy of the stable antisymmetric soliton is four times as large as
that for the fundamental DM soliton with the same width; as is well known,
the \textquotedblleft heavier" soliton provides for better suppression of
the timing jitter, so the antisymmetric one has advantage, in this respect
for applications to fiber-optic telecommunications (see further details in
section 4). Third, the VA predictions are completely verified by direct
simulations, see below.

\subsection{Interactions between antisymmetric solitons}

To describe the interaction between pulses belonging to different channels,
we start with a more general expression for the pulse, which is obtained
from the one-channel solution by the Galilean boost, 
\begin{eqnarray}
u(z,\tau ) &=&u_{0}\left( z,\tau -T_{u}(z)\right) \exp \left( -i\omega
_{u}(\tau -T_{u})+i\psi _{u}(z)\right) ,  \nonumber \\
v(z,\tau ) &=&v_{0}\left( z,\tau -T_{v}(z)\right) \exp \left( -i\omega
_{v}(\tau -T_{v})+i\psi _{v}(z)\right) .  \label{boost}
\end{eqnarray}
Here $T_{u,v}$, $\psi _{u,v}$ and $\omega _{u,v}$ are, respectively, the
position, phase, and frequency shifts. In terms of the Galilean boost, the
latter are constant, while the position shifts evolve in $z$ according to
the equations 
\begin{equation}
\frac{d\,T_{u,v}}{d\,z}=\pm c-\left( D(z)+\epsilon \overline{D}_{u,v}\right)
\omega _{u,v},  \label{temporal}
\end{equation}
which also include a contribution from the group-velocity difference between
the channels.

The application of the VA technique to the two pulses defined as in Eq. 
(\ref{az1}) leads to the following results: the two energies $E_{u,v}\equiv
(1/4)A_{u,v}^{2}W_{u,v}^{3}$ are conserved separately in the channels, and
the other variational parameters evolve according to the following ODEs 
\begin{equation}
\frac{d\,W_{u,v}}{d\,z}=2\left[ D(z)+\epsilon \overline{D}_{u,v}\right]
W_{u,v}b_{u,v};  \label{Wuva}
\end{equation}
\begin{equation}
\frac{d\,\omega _{u,v}}{d\,z}=\frac{\pm 32\epsilon E_{v,u}(T_{u}-T_{v})} 
{\left( W_{u}^{2}+W_{v}^{2}\right) ^{5/2}}\exp \left[ -\frac{
2(T_{u}-T_{v})^{2}}{W_{u}^{2}+W_{v}^{2}}\right] B,  \label{ouv}
\end{equation}
here we define 
\begin{eqnarray}
B &=&\frac{2\left( W_{u}^{2}-W_{v}^{2}\right) ^{2}-7W_{u}^{2}W_{v}^{2}} 
{4\left( W_{u}^{2}+W_{v}^{2}\right) }-\frac{\left( W_{u}^{2}-W_{v}^{2}\right)
^{2}-6W_{u}^{2}W_{v}^{2}}{\left( W_{u}^{2}+W_{v}^{2}\right) ^{2}}
(T_{u}-T_{v})^{2}  \nonumber \\
&&-\frac{4W_{u}^{2}W_{v}^{2}}{\left( W_{u}^{2}+W_{v}^{2}\right) ^{3}}
(T_{u}-T_{v})^{4}.  \nonumber
\end{eqnarray}
Note that, as it follows from Eq. (\ref{ouv}), 
\begin{equation}
\frac{d}{d\,z}\left[ E_{u}\omega _{u}+E_{v}\omega _{v}\right] =0,
\label{FPa}
\end{equation}
which implies the conservation of the net momentum, $P\equiv E_{u}\omega
_{u}+E_{v}\omega _{v}$.

Now, we focus on the most interesting case when the pulse in each channel is
a stable antisymmetric soliton of the same width, i.e., in the absence of
the interaction between them, their parameters are selected according to
Eqs. (\ref{del}) and (\ref{tau}). Then the interaction between the two
antisymmetric solitons is described by a system including a difference of
two equations Eq. (\ref{temporal}), 
\begin{equation}
\frac{d}{dz}\left( T_{u}-T_{v}\right) =2\overline{c}-D(z)\left( \omega
_{u}-\omega _{v}\right) \,-\epsilon \left( \overline{D}_{u}\omega _{u}- 
\overline{D}_{v}\omega _{v}\right) .  \label{T-T}
\end{equation}
and Eqs. (\ref{ouv}) that can be now cast in the form

\begin{equation}
\frac{d\omega _{u,v}}{dz}=\frac{\pm 4\sqrt{2}\epsilon E_{v,u}\tau
_{0}^{3}\left( T_{u}-T_{v}\right) }{\left[ \tau _{0}^{4}+4\Delta ^{2}(z)
\right] ^{3/2}}\exp \left( -\frac{\left( T_{u}-T_{v}\right) ^{2} \tau
_{0}^{2}}{\tau _{0}^{4}+4\Delta ^{2}(z)}\right) C\,,  \label{omega}
\end{equation}
\[
C\equiv -\frac{7}{8}+\frac{3\tau _{0}^{2}\left( T_{u}-T_{v}\right) ^{2}} 
{2\left[ \tau _{0}^{4}+4\Delta ^{2}(z)\right] }-\frac{\tau _{0}^{4}\left(
T_{u}-T_{v}\right) ^{4}}{2\left[ \tau _{0}^{4}+4\Delta ^{2}(z)\right] ^{2}} 
\]

Equations (\ref{omega}) and (\ref{T-T}) constitute a dynamical system
describing the interaction between the antisymmetric solitons and formation
of possible bound states between them, similar to how it was investigated
for fundamental DM solitons in Ref. \cite{we}. Note that the energies 
$E_{u,v}$ do not appear in these equations as arbitrary parameters; instead,
they must be expressed in terms of $\tau _{0}$ and $\overline{D}_{u,v}$ by
means of Eqs. (\ref{del}) and (\ref{tau}) . Arbitrary parameters are $\tau
_{0}$, or the DM strength $S$, the inverse-group-velocity-difference 
$\overline{c}$, and the PADs $\overline{D}_{u,v}$.

The third-order system of Eqs. (\ref{omega}) and (\ref{T-T}) can be further
reduced to a second-order one in the symmetric case, with $\overline{D}_{u}= 
\overline{D}_{v}$ [hence also $E_{u}=E_{v}$, see Eqs. (\ref{del}),(\ref{tau}
)]. Then, defining $T\equiv T_{u}-T_{v}$, $\omega \equiv \omega _{u}-\omega
_{v}$, and $E_{u}=E_{v}\equiv E$, $\overline{D}_{u}=\overline{D}_{v}\equiv 
D$, the reduced system is 
\begin{eqnarray}
\frac{d\omega }{dz} &=&\frac{8\sqrt{2}\epsilon E\tau _{0}^{3}T}{\left[ \tau
_{0}^{4}+4\Delta ^{2}(z)\right] ^{3/2}}\exp \left( -\frac{\tau _{0}^{2}T^{2} 
}{\tau _{0}^{4}+4\Delta ^{2}(z)}\right) C\,,  \label{sym1} \\
\frac{dT}{dz} &=&2\overline{c}-\left[ D(z)+\epsilon \overline{D}\right]
\omega .  \label{sym2}
\end{eqnarray}

It should be pointed out here that, if PAD is zero, it may be necessary to
add the third-order dispersion (TOD) to the DM model, in the case when the
solitons are taken very narrow (in the temporal domain), to provide for a
very high bit rate per channel. Effects of TOD on fundamental DM solitons
have been systematically studied in \cite{TOD}. It was shown in that the TOD
gives rise to an asymmetry of the DM-soliton's profile and generation of
radiation. We anticipate that the effects of TOD on antisymmetric solitons
will be similar. However, detailed investigation on this issue is definitely
beyond the scope of the present paper, being a subject for a separate work.

\section{Comparison with the results of direct simulations}

It is necessary to check the VA equations derived in previous section
against direct simulations. To this end, we solved the underlying equations 
(\ref{1-ch}) and (\ref{2-cha}), (\ref{2-chb}) by a symmetrized split-step
Fourier method, in which the linear part is computed exactly via the fast
Fourier transformation (FFT), and the nonlinear part is evaluated implicitly
via an iteration procedure at the midpoint of the stepsize (see, e.g., Ref. 
\cite{we}).

\subsection{Single ASDM soliton}

First of all, the validity of Eqs. (\ref{del}) and (\ref{tau}), which
predict equilibrium values of the parameters for the ASDM soliton, Eq. 
(\ref{1-ch}) was solved numerically with the parameters of the initial
antisymmetric pulse taken as predicted by these expressions. We fixed 
$L_{1}=L_{2}=0.5$, and $D_{1}=2.0$, $D_{2}=-2.0$, unless specified otherwise.

In the zero-PAD case, direct simulations show that the width parameter of
the pulse is always kept close to $\tau _{0}^{2}=0.301$ regardless of the
value of energy $E$. An example is shown in Fig. 1 for $E=1.0$ and $E=2.0$.
It can be seen that the ASDM soliton remains stable, keeping the same width
as the initial pulse at the end of each dispersion segment.

In the anomalous-PAD case, results of direct simulation agree with the VA
predictions as well. Figure 2 shows the evolution of the antisymmetric
solitons for $\overline{D}=0.05$, and $E=2.0,4.0$. The corresponding widths
are $\tau _{0}^{2}=0.6993$ and $0.5256$, respectively.

In the normal-PAD case, the VA predicts that the antisymmetric DM soliton is
stable only when $|\overline{D}|/E\leq 0.0032$. To test this, we took, for
instance, $\overline{D}=-0.01$ and $E=2.0,4.0$. The soliton is anticipated
to be unstable for $E=2.0$, since $|\overline{D}|/E=0.005$ in this case, and
stable for $E=4.0$, as then $|\overline{D}|/E=0.0025$. These predictions are
confirmed by direct simulations, whose results are shown in Fig. 3 (solid
and dashed curves showing, respectively, the wave profile at $z=800$ and the
initial one). It is seen that, for $E=2.0$, the two parts of the soliton
separate from each other as $z$ increase, but for $E=4.0$ the amplitudes of
the two peaks and the difference between them keep almost the same values as
they had in the initial pulse, even as $z$ takes attains the large value of 
$800$, which is only violated by some radiation loss.

To summarize these results, in Fig. 4 we plot the ratio of the PAD to the
pulse energy, $\overline{D}/E$, versus the pulse's width $\tau _{0}^{2}$ for
the cases of the zero, anomalous, and normal PAD. The solid curves depict
the VA predictions, while the circles are data produced by direct
simulations. Good agreement between the VA and numerical results is obvious.

\subsection{Interactions between antisymmetric solitons and formation of
bound states}

Proceeding to interactions between the antisymmetric solitons, we first
simulated head-on complete collisions, in which case the pulses, moving with
opposite velocities, are well separated before and after the collision.
Basically, the collision features the generic property of the soliton
collision, that is, the pulses pass through each other with position shifts.
However, because the DM pulses considered here are not solitons in the
strict mathematical sense, each one gets slightly distorted by the
interaction, its humps changing their height. A typical example is displayed
in Fig. 5 for $\overline{D}_{u}=\overline{D}_{v}=0.05$, $2\overline{c}=0.1$,
and $E_{u}=E_{v}=2.0$. 
In the direct simulation, it is observed that the two pulses repel each
other at an early stage of the interaction, and attract at a late stage. A
position shift of $\delta T=0.625$ has resulted from the complete collision.
This result agree with the VA results shown in Fig. 5(c), which predict the
position shift $0.65$ and the zero frequency shift for each pulse. The
position shift, along with a possible frequency shift may be considered as
the source of the timing jitter induced by the collisions, which will be
considered in more detail in the next section.

It is predicted by VA that there is no possibility for the formation of BS's
in the case of complete collision. As to the case of incomplete collision,
similar to the possibility of formation of bound states (BS's) of two
fundamental DM soliton belonging to different channels that was found in
Ref. \cite{we}, BS's of the ASDM solitons can be formed too. The difference
is that more energy is needed for the formation of BSs in the latter case.
Figure 6 shows an example for $\overline{D}_{u}=\overline{D}_{v}=0.075$, 
$2\overline{c}=0.05$, $E_{u}=E_{v}=3.0$, two antisymmetric solitons being
initially set at the same position. Since the formation of BSs is
detrimental for the fiber-optic telecommunication systems, the smaller
chance for this effect in the case of the antisymmetric solitons is an
advantage offered by them.

For incomplete collisions in the symmetric situation, with $D_{u}=D_{v}=D$, 
$E_{u}=E_{v}=E$, and $D/E=0.025$, a plot of the minimum energy $E_{\mathrm{min
}}$, necessary for the formation of the BS, vs. $2\overline{c}$, as
predicted by VA is displayed in Fig. 7. The variational predictions are
checked, at several points, against direct simulations, the corresponding
data being marked by rhombuses. As is seen, the agreement between VA and
direct results is good. It is noted that both the VA predictions and direct
simulations yield a critical value of $2\overline{c}_{r}\approx 0.2$, above
which no BSs exist, no matter how large the energy is. In other words, the
formation of a BS is prevented for the values of IGVD exceeding 
$2\overline{c}_{r}$.

However, in some cases the ASDM solitons may be completely distorted by the
interaction, see an example in Fig. 8 for $\overline{D}_{u}=\overline{D}
_{v}=0$, $2\overline{c}=0.1$, and $E_{u}=E_{v}=4.0$. This phenomenon often
happens when the energy is large or the group-velocity difference between
the channels is small, which are detrimental features for the applications.

We also simulated collisions between fundamental and antisymmetric solitons,
see a typical example in Fig. 9. It is seen that both the fundamental and
antisymmetric solitons do not change their shapes after the collision. A
theoretical study of interactions between the fundamental and antisymmetric
DM solitons could be a natural extension of the present work.

\section{Timing jitter of antisymmetric solitons}

\subsection{Estimate of the timing-jitter suppression in one channel}

Based on the variational results for the antisymmetric solitons presented
above, we now aim to estimate the Gordon-Haus timing jitter (generated by
optical noise in the fiber link \cite{Agrawal}) for pulses of this type. We
will follow the procedure of evaluating the jitter which was implemented for
fundamental DM soliton in Ref. \cite{we2}. To this end, we use a known
expression for the jitter-suppression factor ($\mathrm{JSF}$) for the DM
soliton vis-a-vis its NLS counterpart, provided the two have equal energies
(see details in Refs. \cite{we2} and \cite{we}): 
\begin{equation}
\mathrm{JSF}=\frac{\left( \int_{-\infty }^{\infty }\tau
^{2}|u_{0}|^{2}\right) _{\mathrm{DM}}}{\left( \int_{-\infty }^{\infty }\tau
^{2}|u_{0}|^{2}\right) _{\mathrm{NLS}}}~.  \label{jsf}
\end{equation}
Using the analytical approximation (\ref{az2}) for the antisymmetric
soliton, we find 
\begin{equation}
\mathrm{JSF}=-\frac{36}{\pi ^{3/2}}\frac{\tau _{0}^{4}+1/3+(\Delta
_{0}+1)^{2}}{{\tau _{0}}^{6}\left[ \ln \left( \sqrt{1+\tau _{0}^{-4}}+\tau
_{0}^{-2}\right) -2\left( \tau _{0}^{4}+1\right) ^{-1/2}\right] }~.
\end{equation}
For comparison, $\mathrm{JSF}$ for the fundamental DM soliton is \cite{we2} 
\[
\mathrm{JSF}=-\frac{3}{\pi ^{3/2}}\frac{{\tau _{0}}^{4}+1/3+(\Delta
_{0}+1)^{2}}{{\tau _{0}}^{6}\left[ \ln \left( \sqrt{1+\tau _{0}^{-4}}+\tau
_{0}^{-2}\right) -2\left( \tau _{0}^{4}+1\right) ^{-1/2}\right] }. 
\]

Another characteristic of the DM solitons is the stretching factor ($\mathrm{
\ SF}$), which is the ratio the maximum and minimum values of its temporal
width, 
\begin{equation}
\mathrm{SF}\equiv \sqrt{\tau _{0}^{4}+\left( 1+(2\Delta _{0}+1)\right) ^{2}}
/\tau _{0}^{2}~.
\end{equation}
A certain compromise between the $\mathrm{JSF}$ and $\mathrm{SF}$ must be
reached in designing a transmission line for DM solitons. In Fig. 10, 
$\mathrm{JSF}$ is plotted versus $\mathrm{SF}$ for both the fundamental
(dashed curve) and antisymmetric (solid curve) DM solitons. It can be seen
that much higher (four times) energy is needed to support the same width of
the antisymmetric soliton, in comparison with the fundamental one. As a
result, the $\mathrm{JSF}$ for the antisymmetric soliton is $12$ times
larger than for the fundamental one. This is a potential advantage for using
ASDM solitons in fiber-optic telecommunications.

\subsection{Collision-induced pulse timing jitter}

One of the serious problems in the use of multi-channel (WDM) schemes is the
crosstalk timing jitter, induced by collisions of pulses belong to different
channels. Here, we aim to estimate the crosstalk jitter induced by
collisions between antisymmetric DM solitons belonging to two adjacent
channels, in the case of both complete and incomplete collisions.

We will follow the approach to this problem developed for the fundamental DM
pulses in Ref. \cite{KMY}. Straightforward use of general expressions for
the collision-induced frequency and position shifts, $\delta \omega _{u,v}$
and $\delta T_{u,v}$ produced by the collision, which were derived in that
work, yields the following results for the ASDM solitons. In the lowest
approximation, $\delta \omega _{u,v}=0$, and 
\begin{equation}
\delta T_{u}=\frac{\sqrt{2\pi }{\epsilon }^{2}\overline{D}_{u}E_{v}}{2{\ 
\overline{c}}^{2}}  \label{T-shift}
\end{equation}
For a typical example corresponding to Fig. 5 (see above), with $\overline{D}
_{u}=\overline{D}_{v}=0.05$, $2\overline{c}=0.1$, $E_{u}=E_{v}=2.0$, Eq. 
(\ref{T-shift}) yields $\delta T_{u}=0.501$. This result agrees well with
that produced by numerical integration of the full VA equations (\ref{sym1})
and (\ref{sym2}), as well as with direct simulations of the underlying
equations (\ref{2-cha}) and (\ref{2-chb}), as is seen in Fig. 5.

For incomplete collisions, in which two pulses are initially overlapped, the
general formulas borrowed from Ref. \cite{KMY} yield the following result
for the largest size of the frequency shift, corresponding to the worst
case, when the solitons begin their interaction at the point where their
centers coincide: 
\begin{equation}
(\delta \omega )_{u}^{\mathrm{max}}=-2\sqrt{2}\epsilon E_{v}\tau _{0}^{3} 
\overline{c}^{-1}\left\langle \left( \tau _{0}^{4}+4\Delta ^{2}\right)
^{-3/2}\right\rangle ,  \label{max}
\end{equation}
with $\left\langle ...\right\rangle $ standing for the average over the DM
period. After evaluating the average value, we obtain from Eq. (\ref{max}), 
\begin{equation}
(\delta \omega )_{u}^{\mathrm{max}}=-2\sqrt{2}\epsilon E_{v}{\tau _{0}}^{-1} 
{\overline{c}}^{-1}\left( \frac{\Delta _{0}+1}{\sqrt{{\tau _{0}}
^{4}+4(\Delta _{0}+1)^{2}}}-\frac{\Delta _{0}}{\sqrt{{\tau _{0}}^{4}+4{
\Delta _{0}}^{2}}} \right) .  \label{max2}
\end{equation}
Assuming that parameters for the antisymmetric solitons are selected as for
the stationary pulses in one channel, i.e., $\Delta _{0}=-1/2$, Eq. 
(\ref{max2}) is simplified to 
\begin{equation}
(\delta \omega )_{u}^{\mathrm{max}}=-\frac{2\sqrt{2}\epsilon E_{v}} 
{\overline{c}\tau _{0}\sqrt{\tau _{0}+1}}~.
\end{equation}
Then, for a large propagation distance $z$, the position shift generated by
the frequency shift grows as $\delta T_{u}^{(\omega )}=-\delta \omega
\epsilon D_{u}z.$

\section{Conclusion}

In this paper, we have studied the propagation and interactions of
antisymmetric solitons in a fiber-optic link subject to strong DM. By means
of the variational approximation (VA), we have obtained analytical
expressions for the initial chirp and width of the antisymmetric pulse at
which the pulse should propagate stably. Interactions between ASDM solitons
belonging to two adjacent channels were also investigated, including the
possibility of the formation of bound states between them. In most cases,
the results predicted by the VA compare quite well with direct simulations
for the underlying partial differential equations. However, in some cases we
the collision between the ASDM solitons may destroy them, which is of course
not predicted by the VA.

We have also estimated the Gordon-Haus timing jitter for the ASDM solitons.
A noteworthy finding is that the jitter-suppression factor for the
antisymmetric solitons may be much larger (by a factor of $12$) than its
previously known counterpart for the fundamental solitons in the same DM
link. The crosstalk jitter, induced by inter-channel collisions between the
antisymmetric solitons in a WDM system, was evaluated too. For complete
collisions, the frequency shift is negligible, whereas the position shift is
significant. Incomplete collisions are most dangerous, generating a finite
frequency shift, which was estimated.

Results reported in this work suggest further investigation in several
directions. In particular, as it was briefly mentioned above, it would be
relevant to study how higher-order effects, such as the TOD and the
intrapulse stimulated Raman scattering, act on the antisymmetric DM
solitons. Interactions between fundamental and antisymmetric DM solitons, as
well as a possibility of formation of bound states between them, may be
another issue to be considered in the future. Lastly, for practical
applications to WDM schemes, it would be useful to study multi-channel
systems, rather than only the dual-channel one.

\newpage

\begin{center}
\textbf{Figure Captions}
\end{center}

Fig. 1. The profiles of the stable ASDM soliton (shown is $\left\vert u(\tau
)\right\vert $), as found from direct simulations of Eq. (\ref{1-ch}) in the
case of the zero path-average dispersion ($\overline{D}=0$), for (a) 
$E=1.0$; (b) $E=2.0$. Dashed curve: $z=0$; solid curve: $z=800$.

Fig. 2. The same as in Fig. 1 (except for that the solid curve pertains to 
$z=400$) in the case of anomalous path-average dispersion, with $D_{0}=0.05$,
and (a) $E=2.0$; (b) $E=4.0$.

Fig. 3. The same as in Fig. 1 in the case of normal path-average dispersion
with $D_{0}=-0.01$, and (a) $E=2.0$ and (b) $E=4.0$, which correspond to the
stable and unstable antisymmetric solitons, respectively.

Fig. 4. The ratio of the PAD to the pulse's energy versus its width in the
cases of the zero, anomalous, and normal path-average dispersion. Solid
curve: VA prediction; circles: numerical results.

Fig. 5. A typical example of the complete collision between two ASDM
solitons with $E_{u}=E_{v}=2.0$, $\overline{D}_{u}=\overline{D}_{v}=0.05$,
and $2\overline{c}=0.1$. (a) The shape of $|u|$ at $z=0$ (dashed curve) and 
$z=400$ (solid curve); (b) the shape of $|v|$ at $z=0$ (dashed curve) and 
$z=400$ (solid curve);
(c) the evolution of $\omega _{u}-\omega _{v}$ (dashed curve) and 
$T_{u}-T_{v}$ (solid curve) as predicted by the VA.

Fig. 6. A bound state of ASDM solitons with $\overline{D}_{u}=\overline{D}
_{v}=0.075$, $2\overline{c}=0.1$ and $E_{u}=E_{v}=3.0$, found from direct
simulations of Eqs. (\ref{2-cha}) and (\ref{2-chb}). The panels (a) and (b)
show the shapes of the bound solitons at $z=400$ (solid curve) and the
initial profile (dashed curve) for $\left\vert u(\tau )\right\vert $ and 
$\left\vert v(\tau )\right\vert $, respectively.

Fig. 7. The minimum energy necessary for the formation of a bound state of
two antisymmetric solitons in the two-channel system vs. the
inverse-group-velocity difference $2\overline{c}$ in the case of incomplete
collisions in a symmetric situation, with $D_{u}=D_{v}=D$, $E_{u}=E_{v}=E$,
and $D/E=0.025$. The minimum energy predicted by VA is shown by the solid
line. Rhombuses represent data points collected from direct simulations.

Fig. 8. An example of destruction of the ASDM solitons as a result of the
collision with $\overline{D}_{u}=\overline{D}_{v}=0.0$, $2\overline{c}=0.1$
and $E_{u}=E_{v}=4.0$, as found from direct simulations of Eqs. 
(\ref{2-cha}) and (\ref{2-chb}). The panels (a) and (b) 
show the shapes of $\left\vert
u(\tau )\right\vert $ and $|v(t)|$ at $z=0$ and $z=400$.

Fig. 9. An example of the collision between fundamental and ASDM solitons
with $E_{u}=E_{v}=2.0$, $\overline{D}_{u}=\overline{D}_{v}=0.05$ and 
$2\overline{c}=0.2$. The panels (a) and (b) show the shapes of $|u|$ and $|v|$
for the fundamental and asymmetric solitons, respectively, at $z=0$ (dashed
curve) and $z=200$ (solid curve).

Fig. 10. The jitter suppression factor versus the stretching factor for the
fundamental (dashed) and antisymmetric (solid) DM solitons.

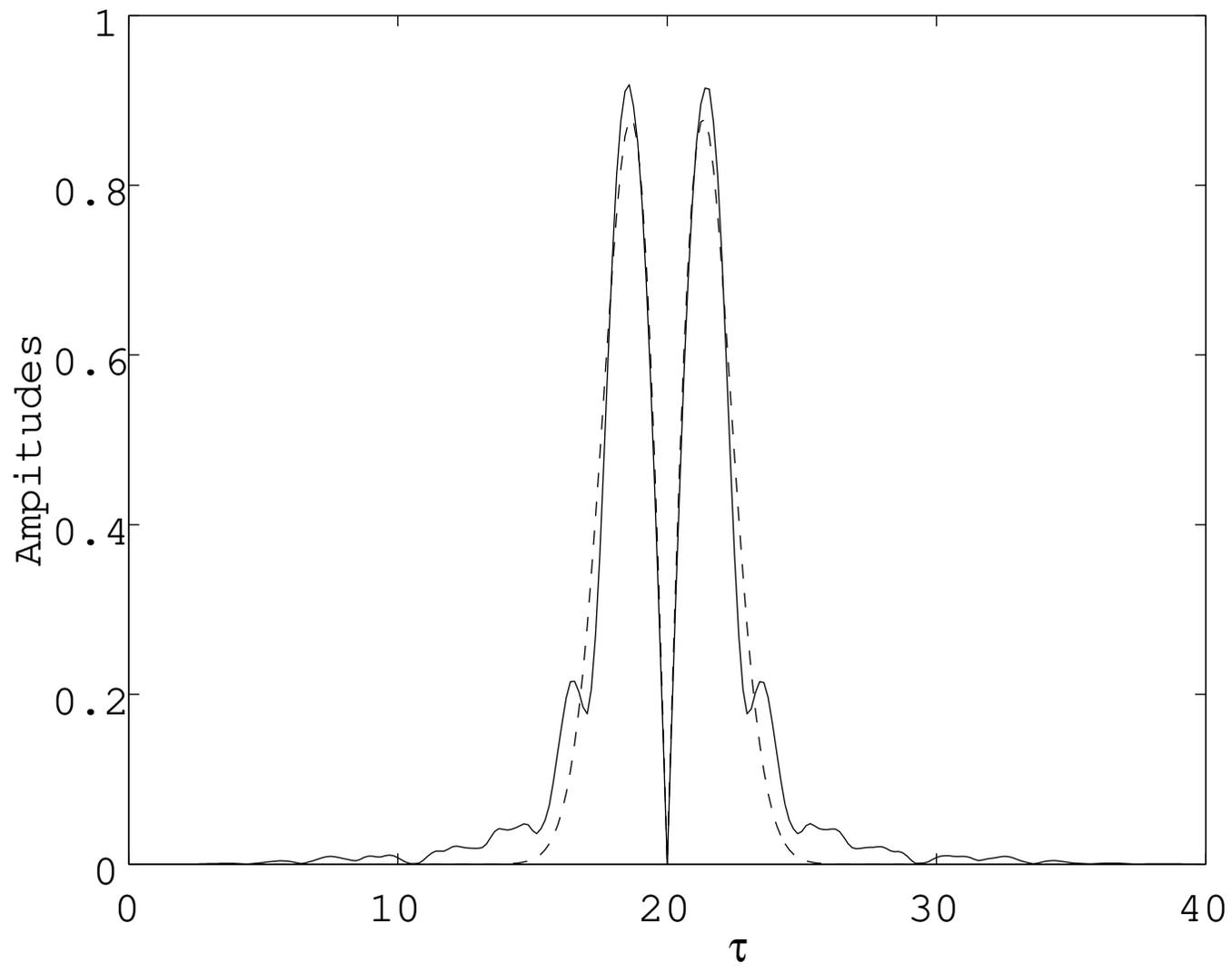
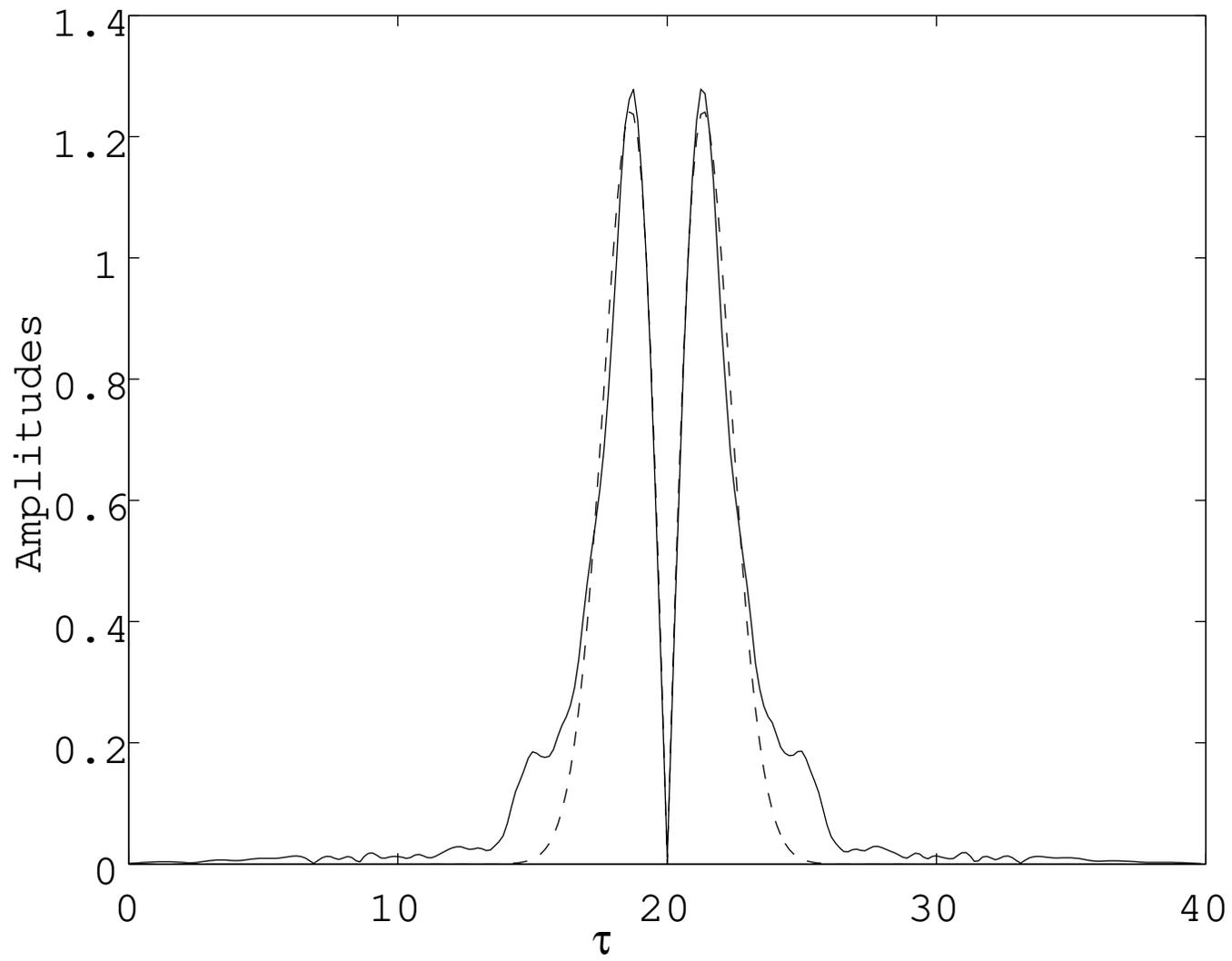
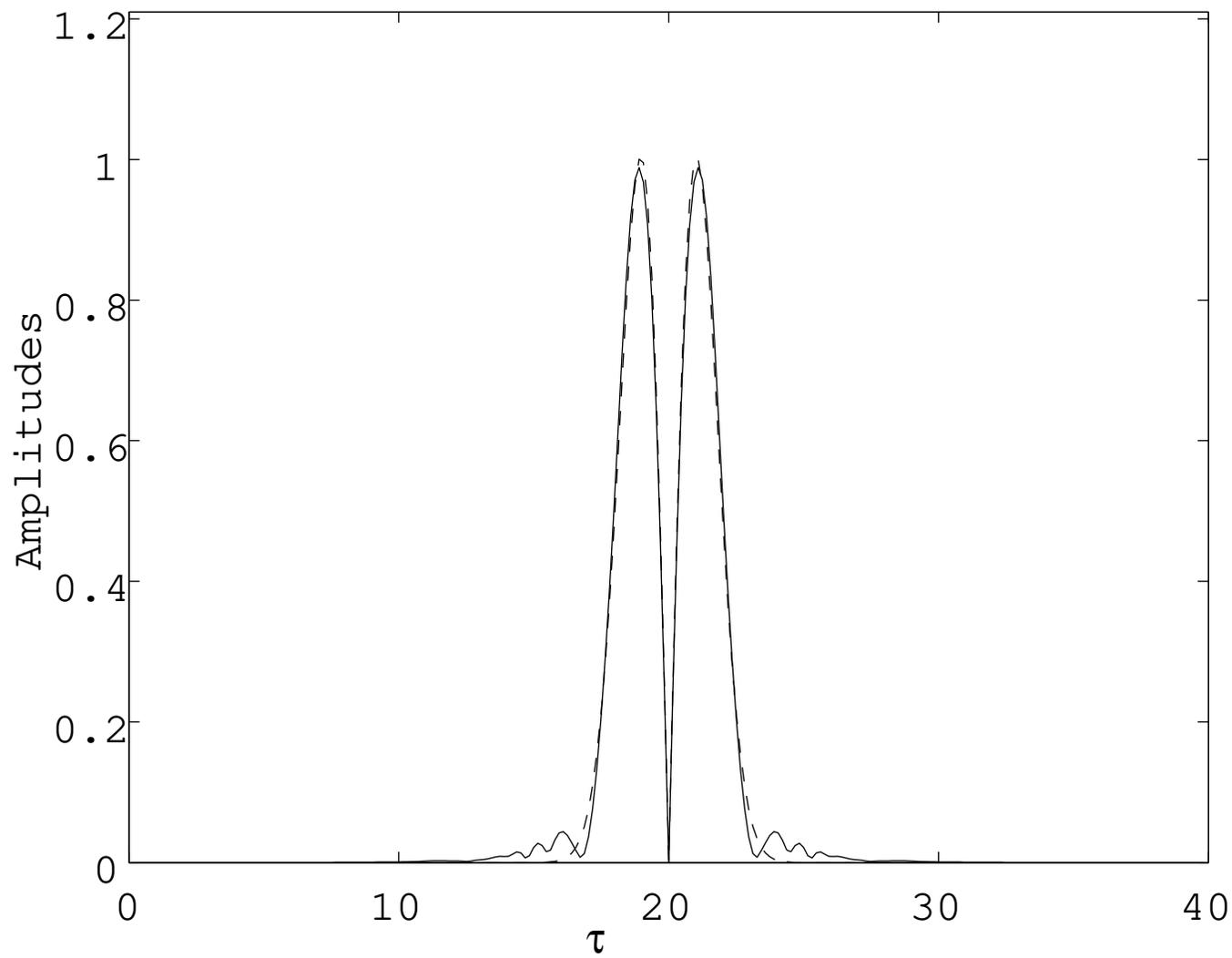
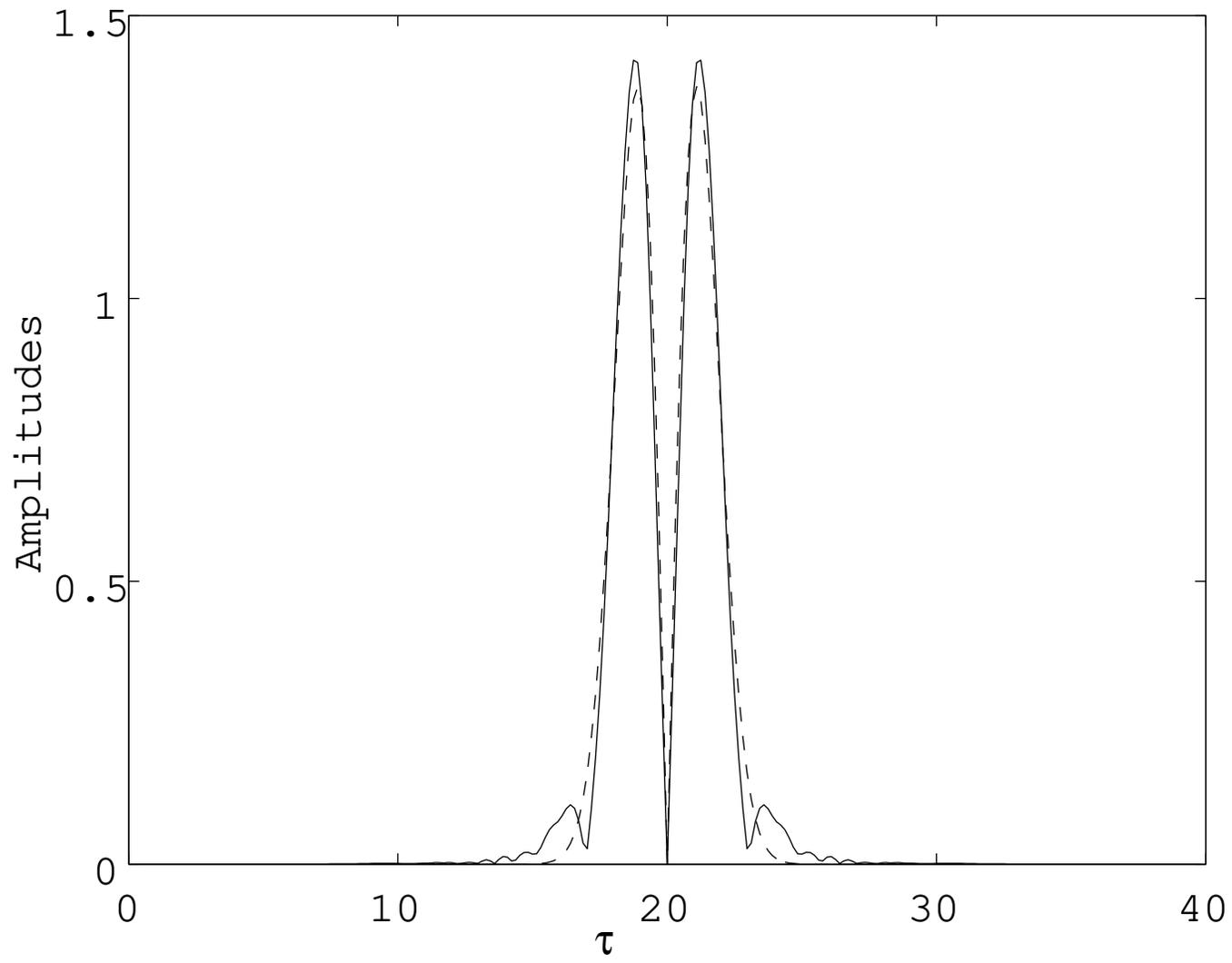
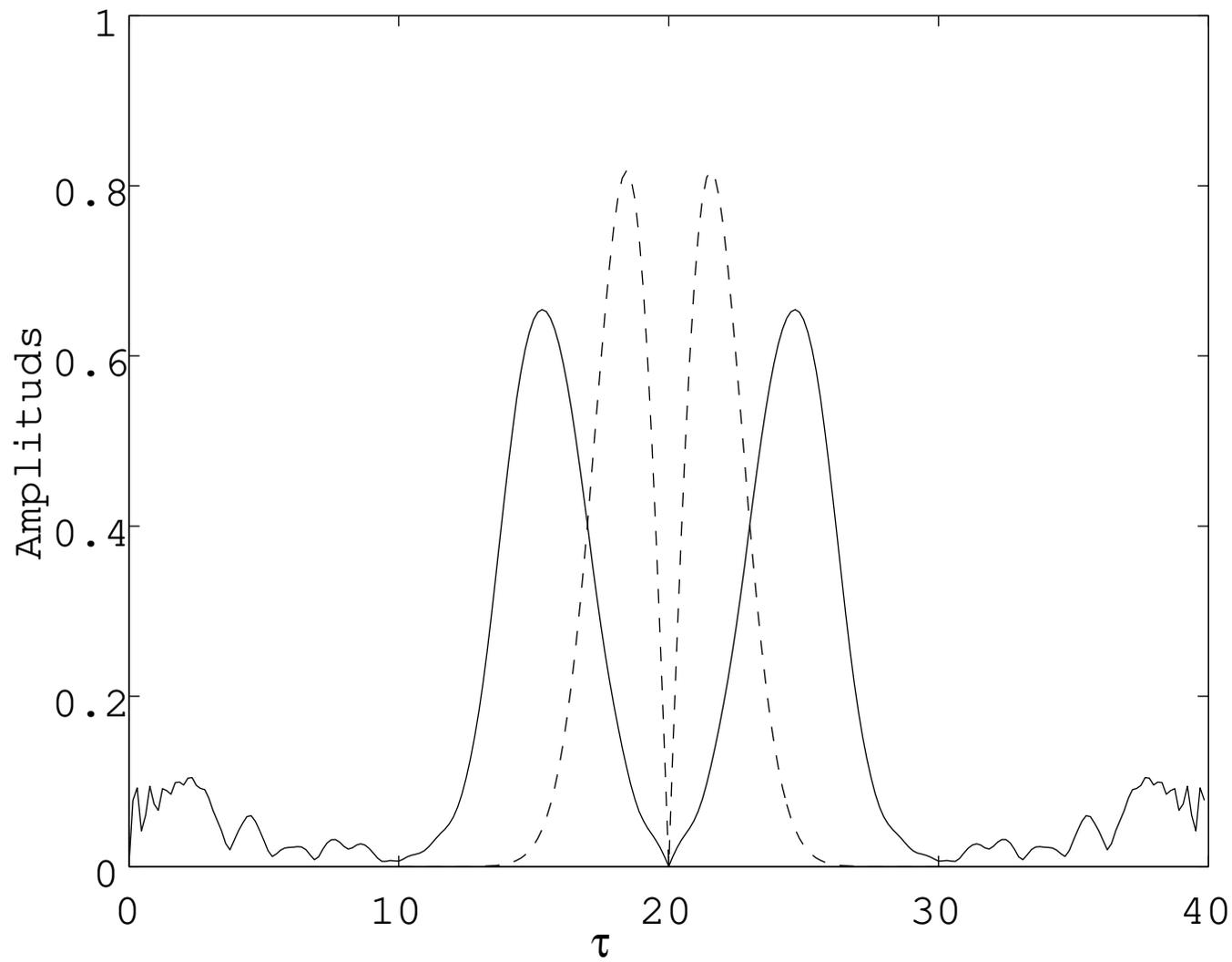
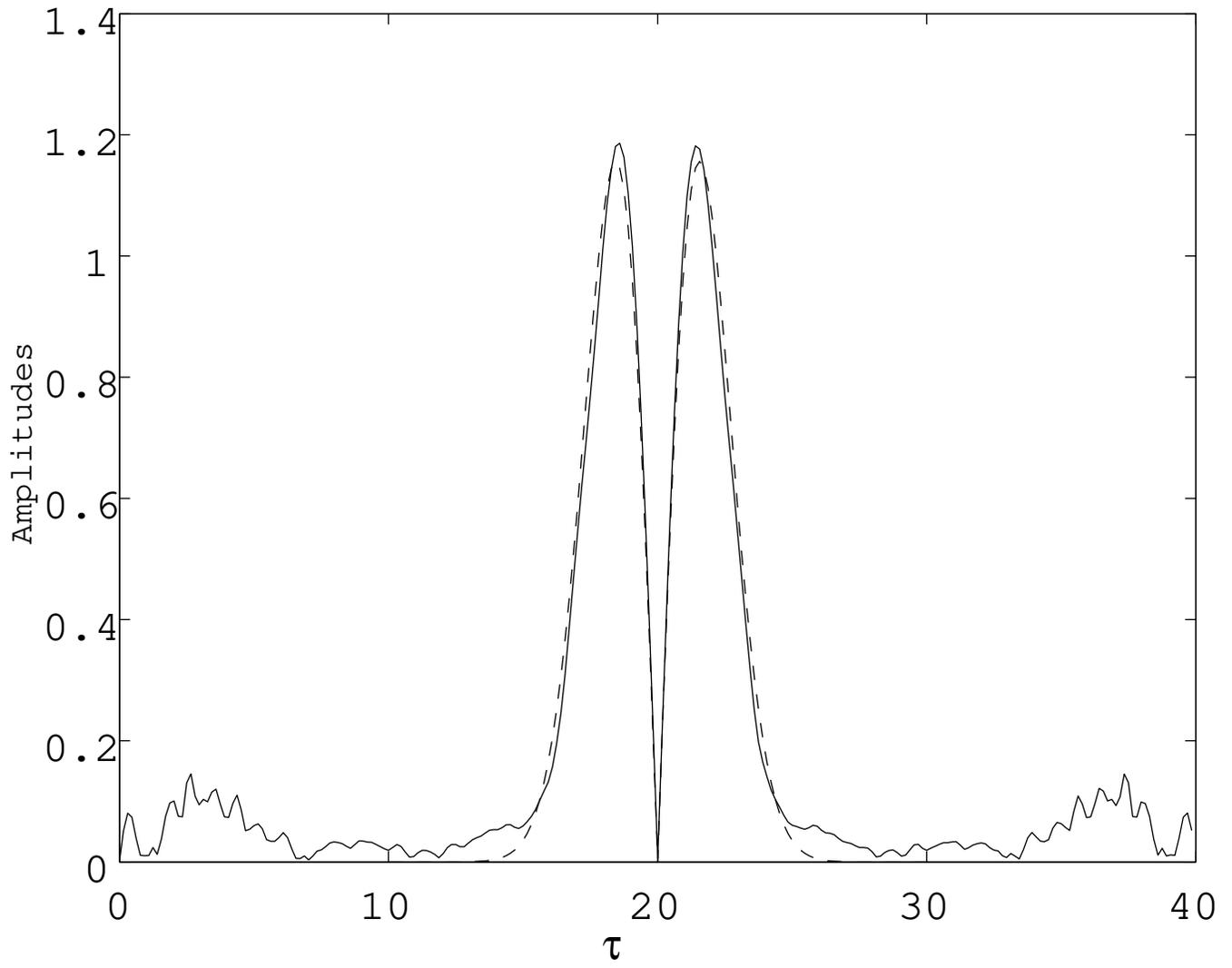
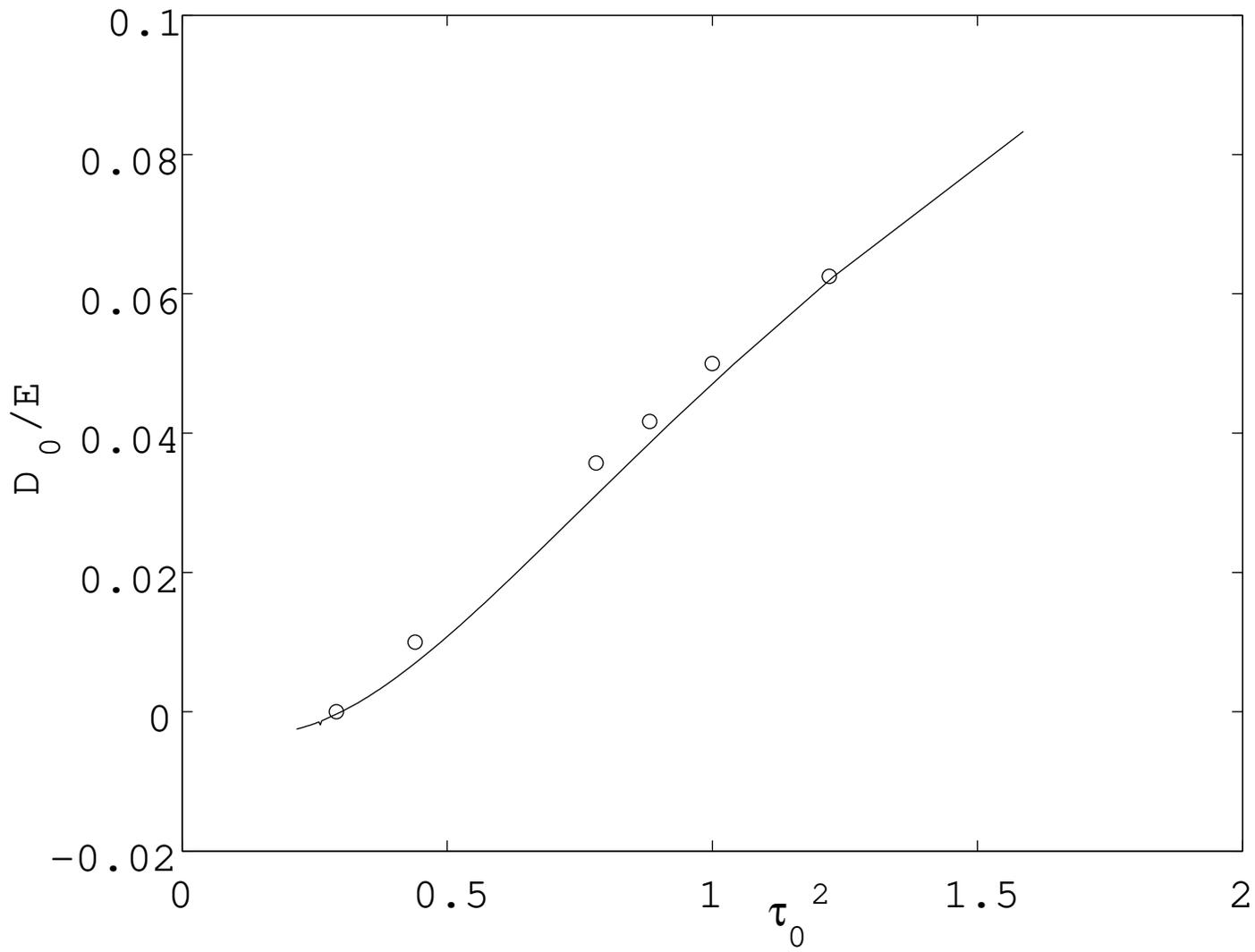
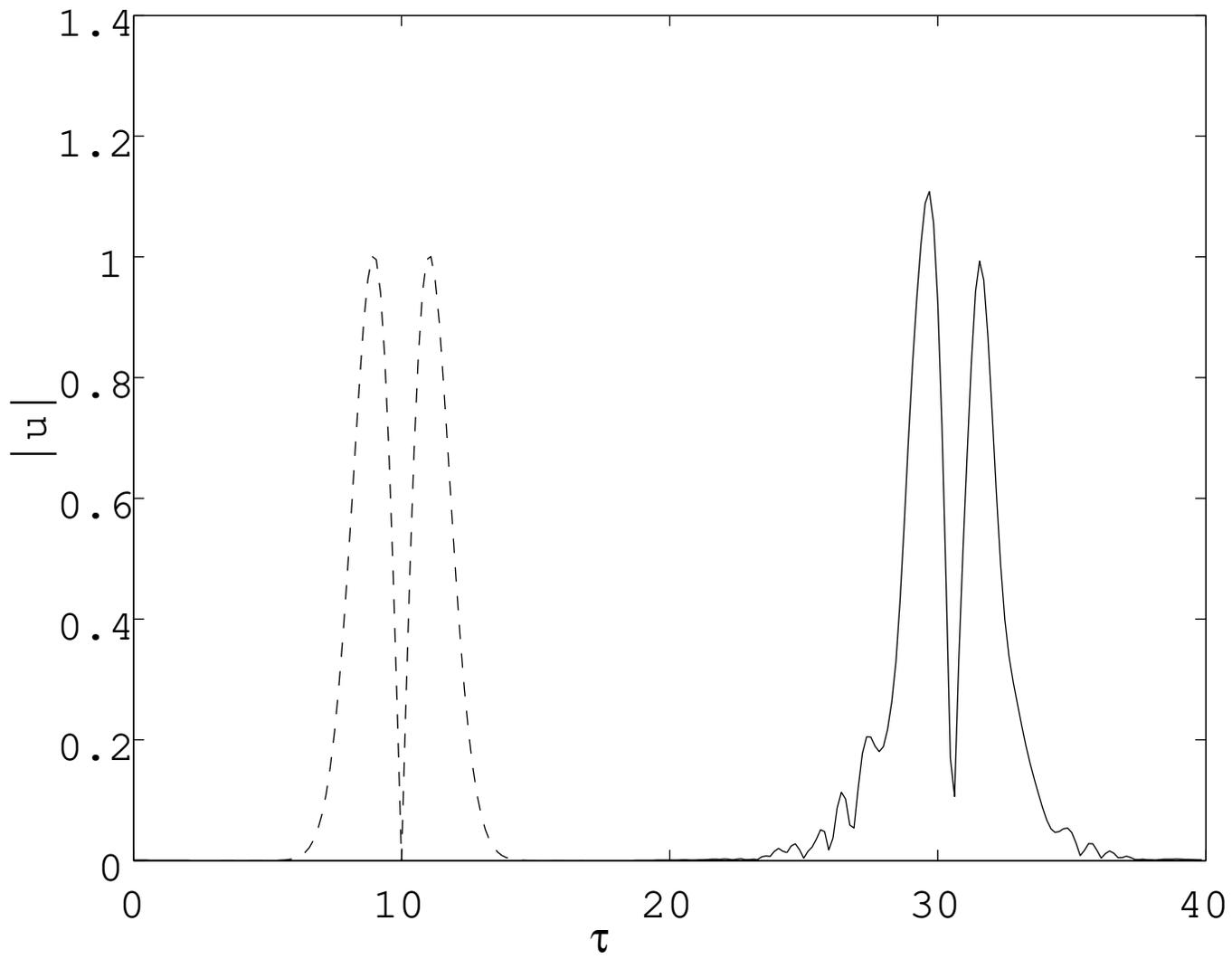
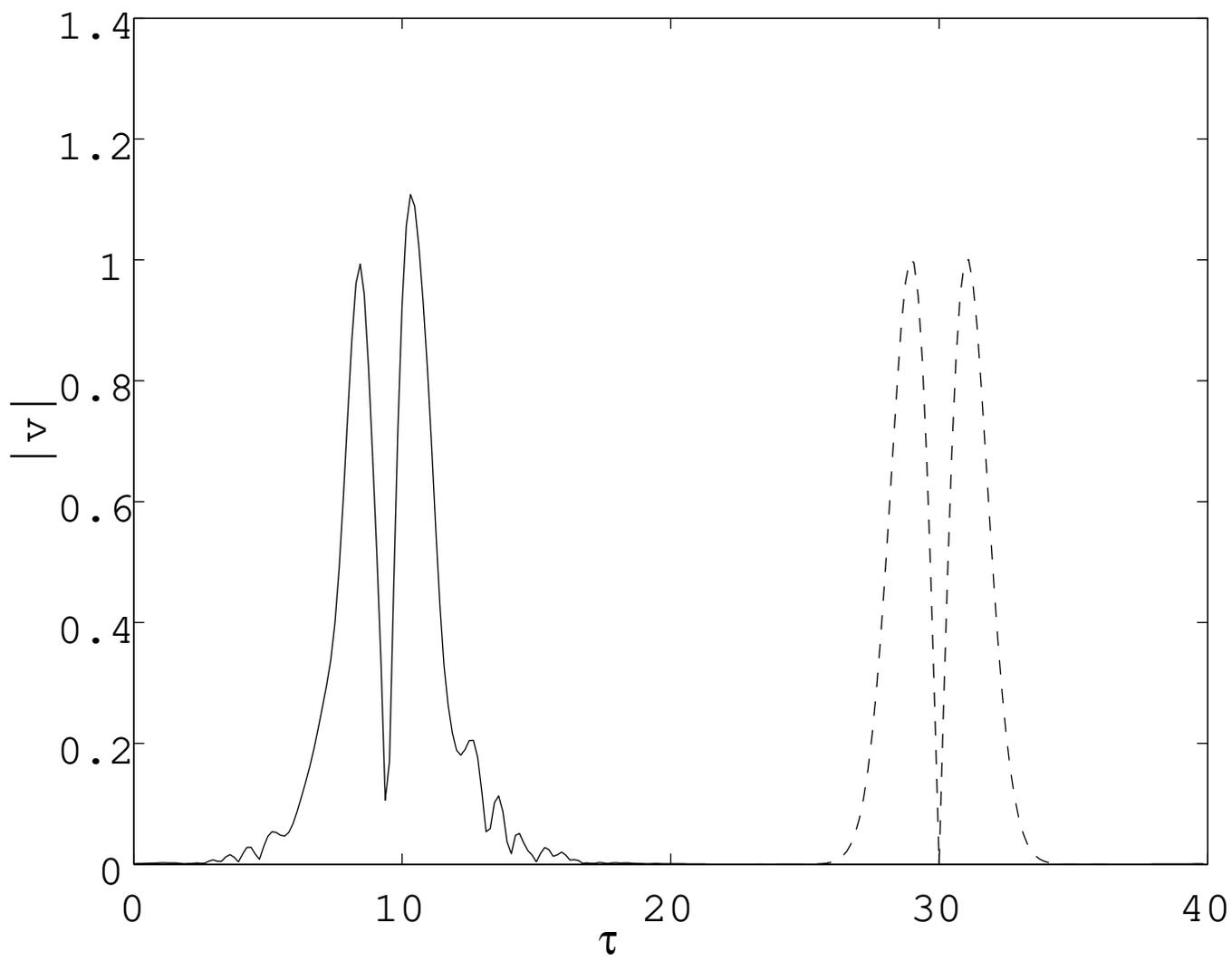
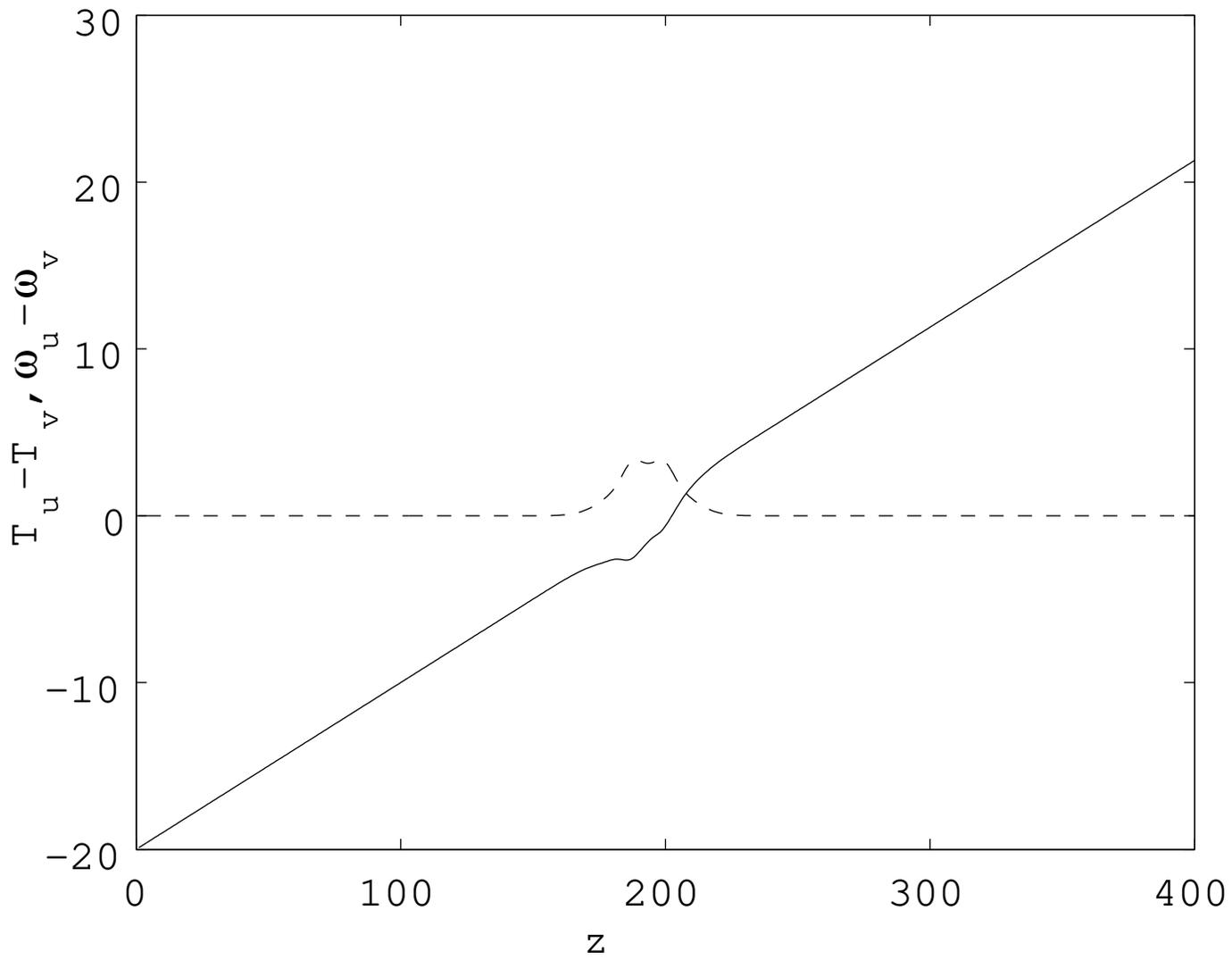
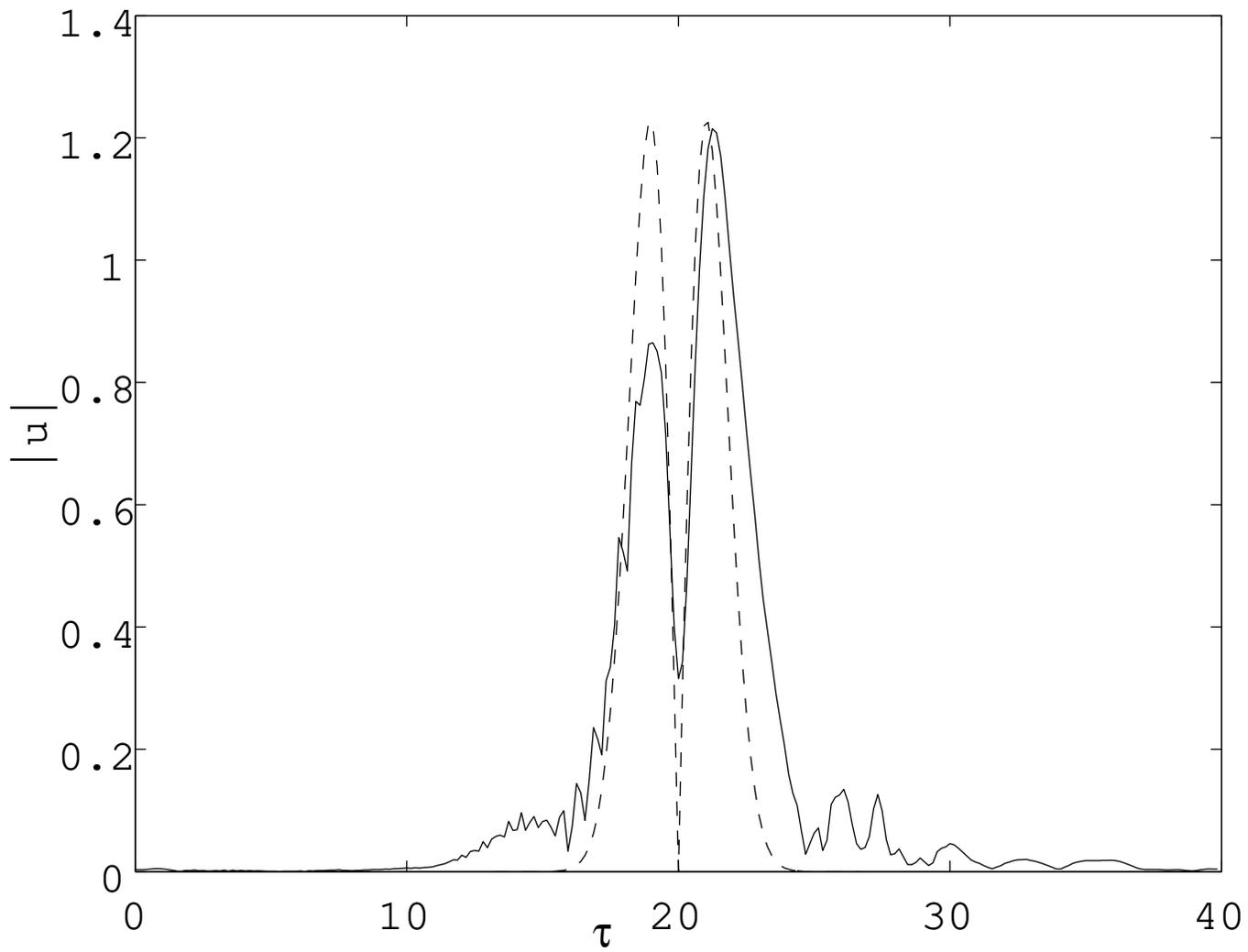
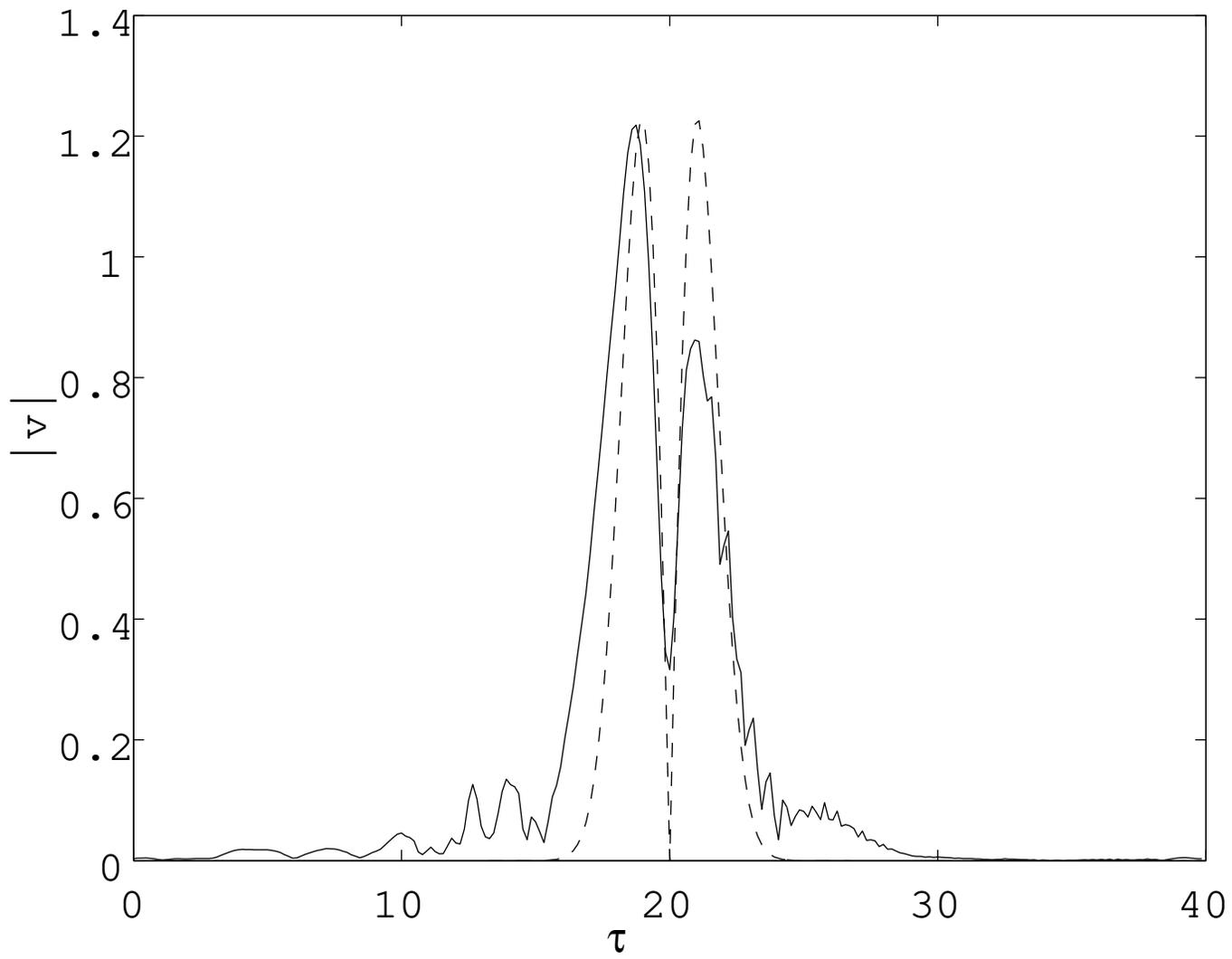
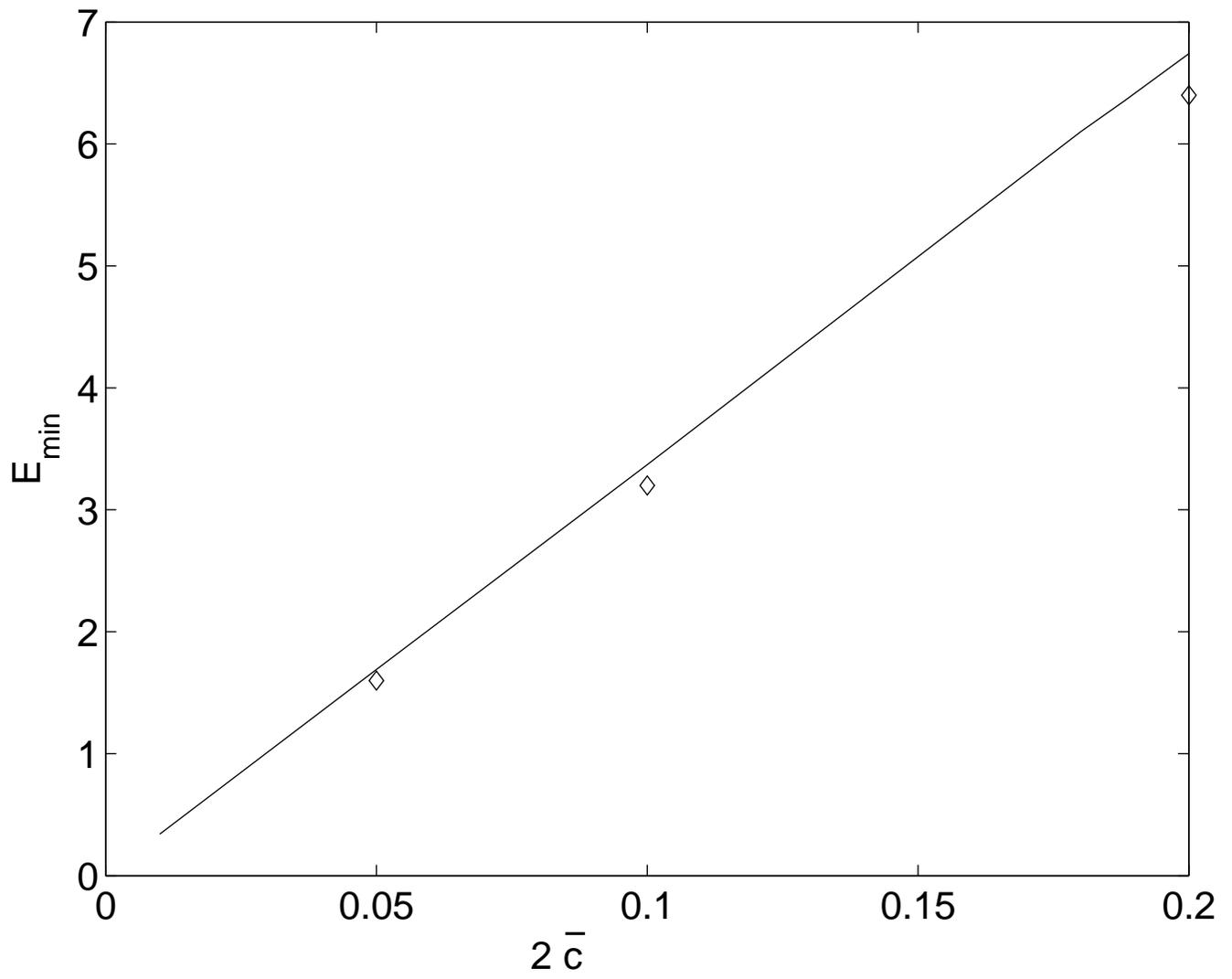
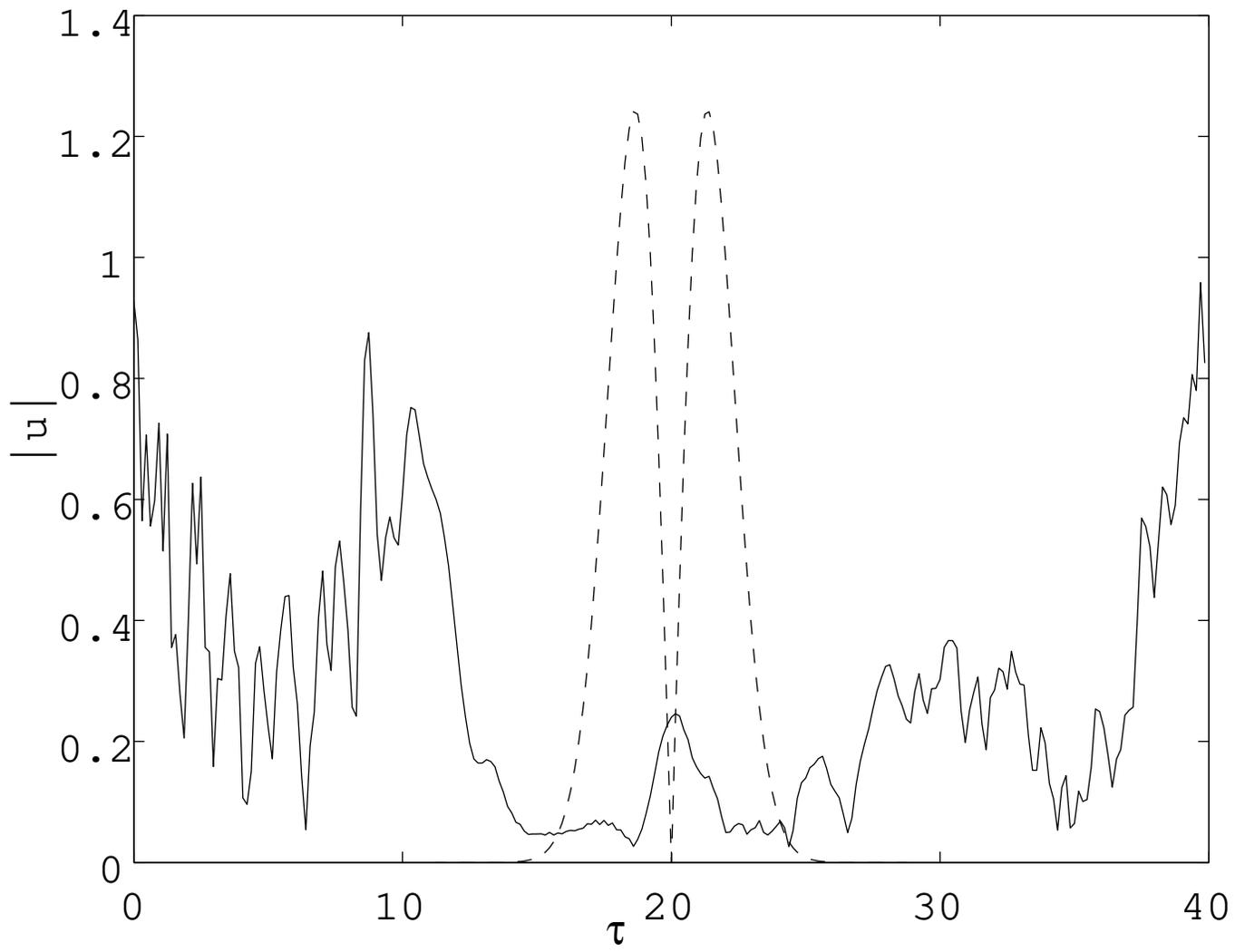
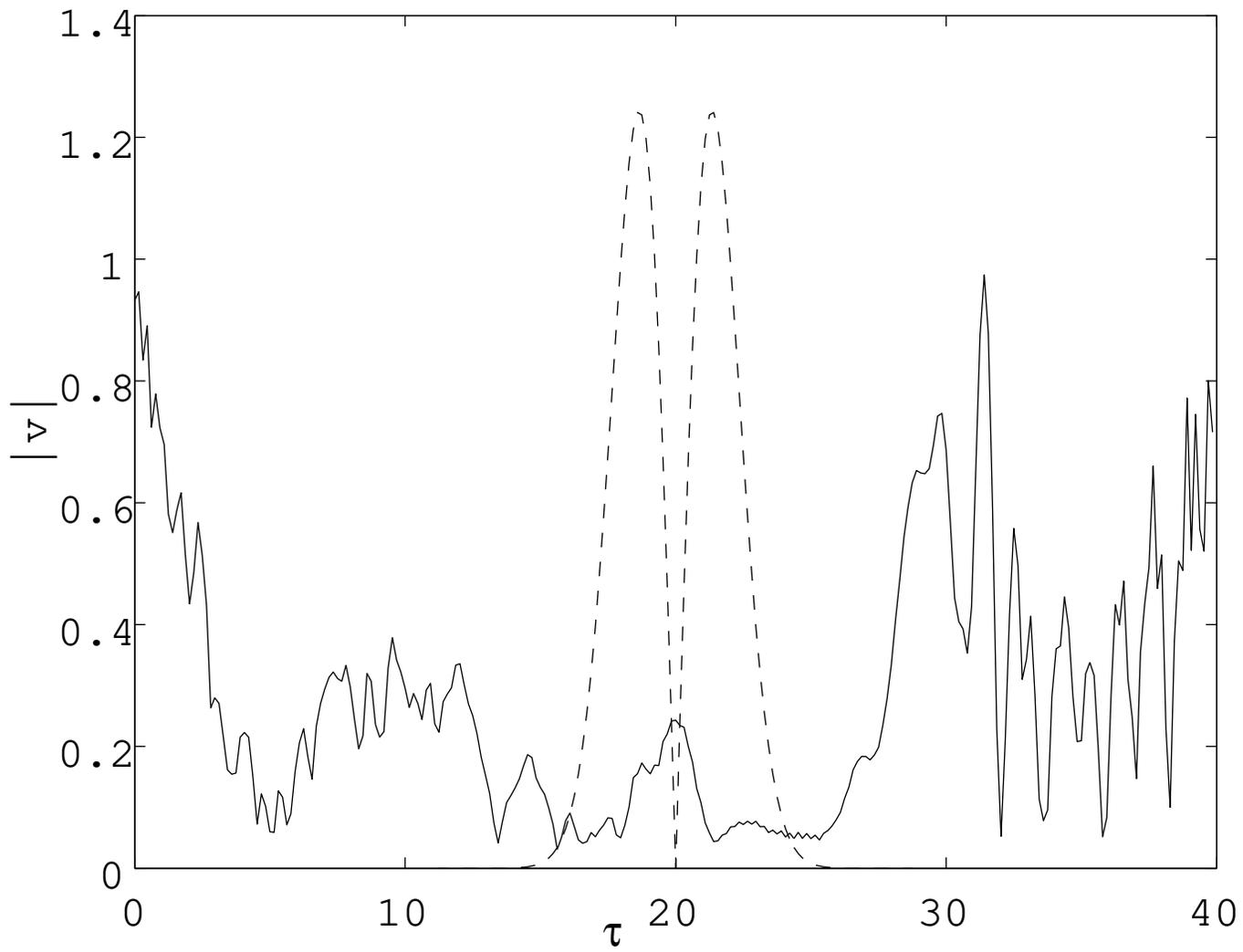
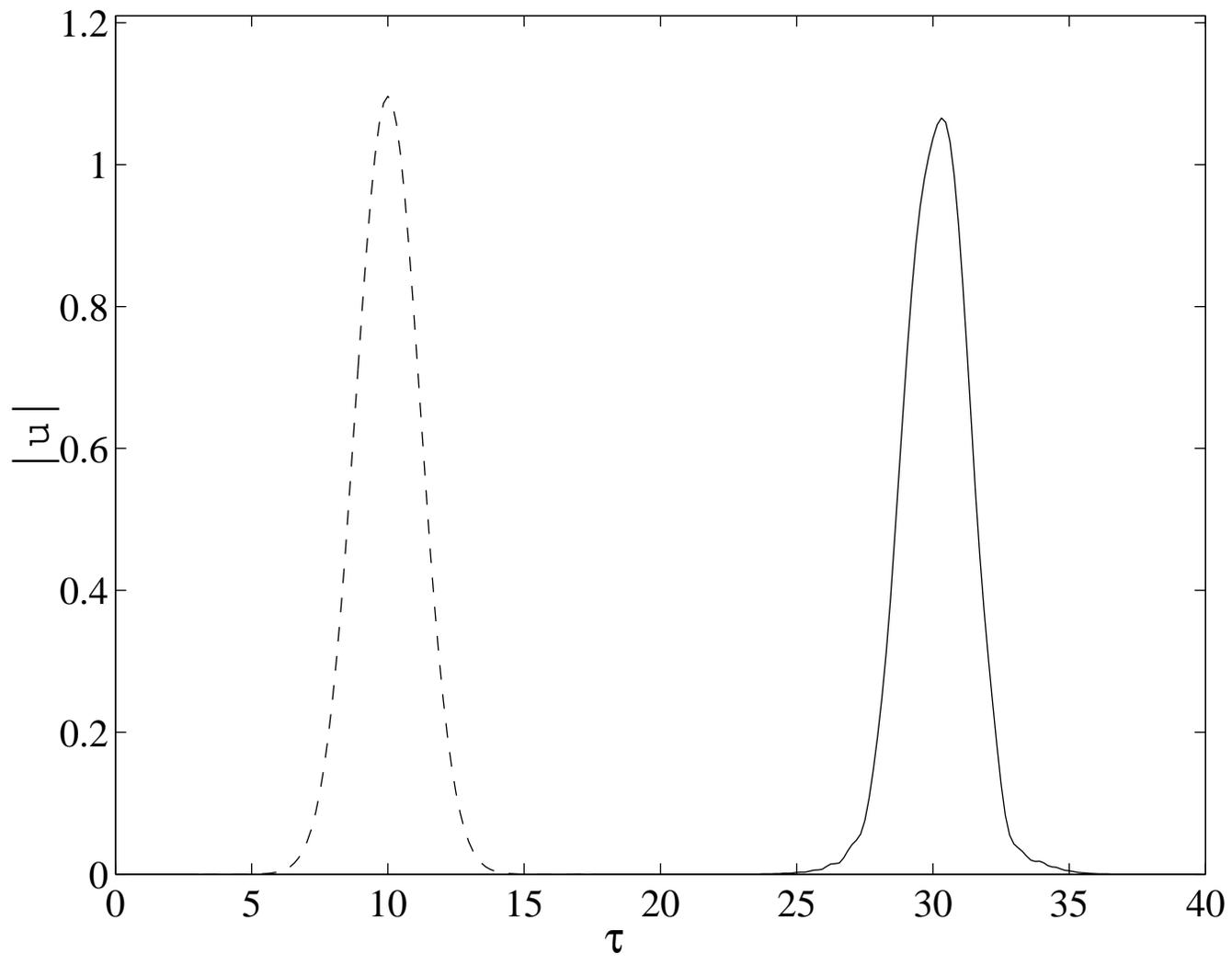
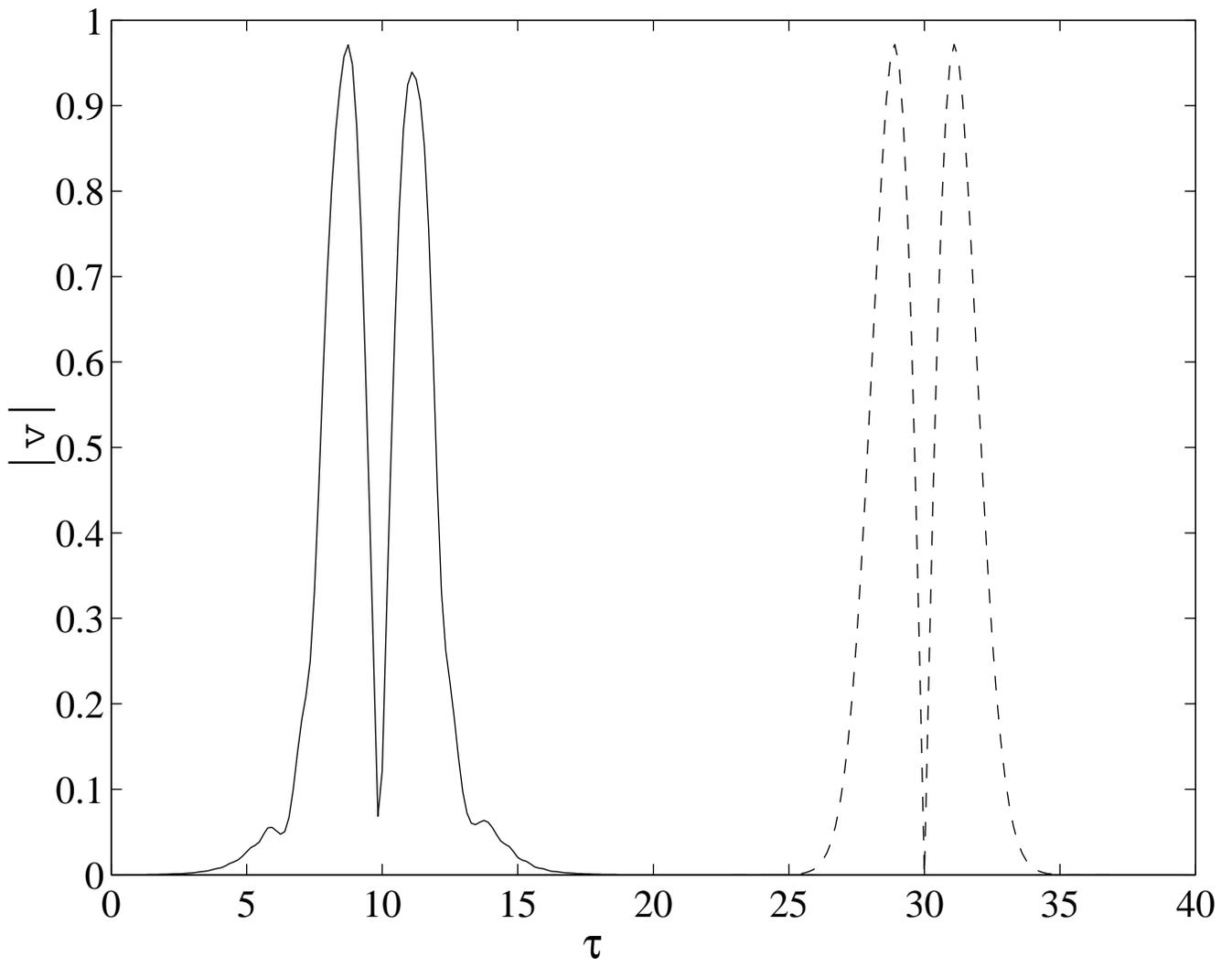
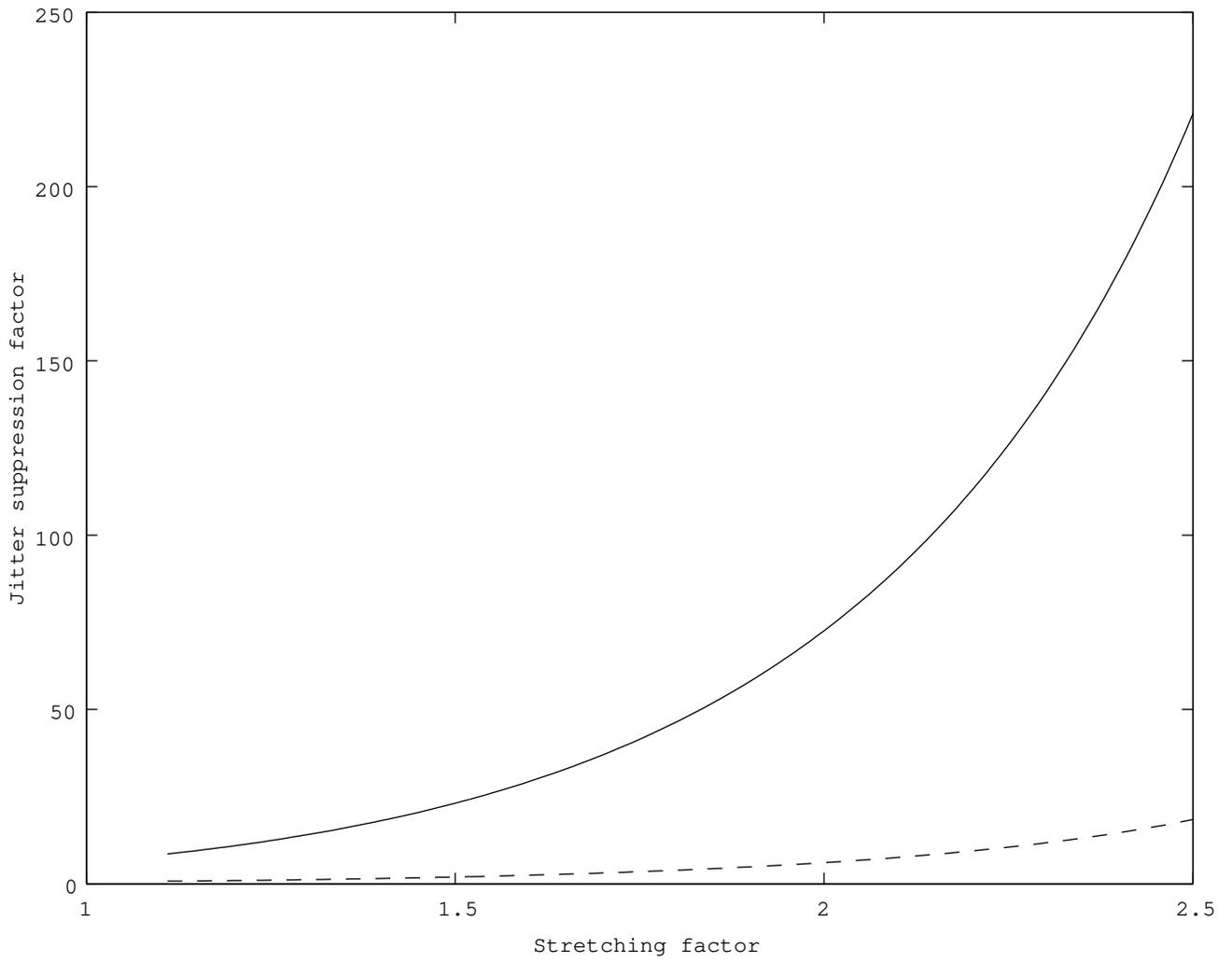
\end{document}